\documentclass[aip,pof]{revtex4-1}

\usepackage[latin1]{inputenc}
\usepackage[english]{babel}
\usepackage{bm}
\usepackage{psfrag}
\usepackage{graphicx}
\usepackage{overpic}
\usepackage{color}
\usepackage{hyperref}

\newcommand\Rey{\mbox{\textit{Re}}} 
\newcommand\eg{e.g.\ }
\newcommand\etal{\mbox{\textit{et al.}}}
\newcommand\be{\begin{equation}}
\newcommand\ee{\end{equation}}

\def\00{\mathbf{0}}
\def\BB{\mathbf{B}}
\def\CC{\mathbf{C}}
\def\s2{G_{opt}^2}
\def\DUs2{\bnabla_{\UU_c} \s2}
\def\FF{\mathbf{F}}
\def\LL{\mathbf{L}}
\def\Pa{P^\dag} 
\def\PP{\mathbf{P}}
\def\QQa{\mathbf{Q}^\dag}
\def\QQ{\mathbf{Q}}
\def\Ua{U^\dag} 	
\def\UUa{\mathbf{U}^\dag}		
\def\UU{\mathbf{U}}
\def\Va{V^\dag} 	
\def\aa{\mathbf{a}}
\def\bb{\mathbf{b}}
\def\ex{\mathbf{e}_x}
\def\ey{\mathbf{e}_y}
\def\bdelta{\boldsymbol{\delta}}
\def\ev{\sigma}
\def\ff{\mathbf{f}}
\def\qq{\mathbf{q}}
\def\nn{\mathbf{n}}
\def\rl{{l_c}}
\def\uu{\mathbf{u}}
\def\nnu{\Rey^{-1}}

\providecommand\bnabla{\boldsymbol{\nabla}}
\providecommand\bcdot{\boldsymbol{\cdot}}

\definecolor{mygreen}{rgb}{0,0.5,0}
\definecolor{myyellow}{rgb}{1,0.5,0} 

\begin{document}


\title{Open-loop control of noise amplification in 
a separated boundary layer flow
} 

\author{E. Boujo}
\email[]{edouard.boujo@epfl.ch}
\affiliation{Laboratory of Fluid Mechanics and Instabilities, 
		\'Ecole Polytechnique F\'ed\'erale de Lausanne,
         CH-1015 Lausanne, Switzerland}

\author{U. Ehrenstein}
\email[]{ehrenstein@irphe.univ-mrs.fr}
\affiliation{Aix Marseille Universit\'e, 
			CNRS, Centrale Marseille, 
			IRPHE UMR 7342, 
			F-13384, Marseille, France}
         
\author{F. Gallaire}
\email[]{francois.gallaire@epfl.ch}
\affiliation{Laboratory of Fluid Mechanics and Instabilities, 
		\'Ecole Polytechnique F\'ed\'erale de Lausanne,
         CH-1015 Lausanne, Switzerland}

\date{December 2013}

\begin{abstract}
Linear optimal gains are computed for the subcritical two-dimensional separated boundary-layer flow past a bump.
Very large optimal gain values are found, making it possible for small-amplitude noise to be strongly amplified and to destabilize the flow.
The optimal forcing is located close to the summit of the bump, while the optimal response is the largest in the shear layer.
The largest amplification occurs at frequencies corresponding to eigenvalues which first become unstable at higher Reynolds number.
Non-linear direct numerical simulations show that a low level of noise is indeed sufficient to trigger random flow unsteadiness, characterized here by large-scale vortex shedding.

Next, a variational technique is used to compute efficiently the sensitivity of optimal gains to steady control (through source of momentum in the flow, or blowing/suction at the wall).
A systematic analysis at several frequencies identifies the bump summit as the most sensitive region for control with wall actuation.
Based on these results, a simple open-loop control strategy is designed, with steady wall suction at the bump summit.
Linear calculations on controlled base flows confirm that optimal gains can be drastically reduced at all frequencies.
Non-linear direct numerical simulations also show that this control allows the flow to withstand a higher level of stochastic noise without becoming non-linearly unstable, thereby postponing bypass transition.

In the supercritical regime, sensitivity analysis of eigenvalues supports the choice of this control design.
Full restabilization of the flow is obtained, as evidenced by direct numerical simulations  and  linear stability analysis. 
\end{abstract}


\maketitle 

\section{Introduction}

Flows can undergo bifurcation well below the critical Reynolds number $\Rey_c$ predicted by linear stability analysis.
Examples of such subcritical flows include both parallel flows (\eg Couette and Hagen-Poiseuille, which are linearly stable for any Reynolds number\cite{Sch01}, i.e. $\Rey_c=\infty$) and non-parallel flows (\eg jets, backward-facing step).
Classical linear stability theory, which focuses on the long-term fate of small perturbations, predicts that all linear eigenmodes are damped for $\Rey < \Rey_c$.
However, it has become clear in the past decades that perturbations can be amplified by  non-modal mechanisms, a phenomenon that  modal linear stability analysis fails to capture \cite{Tre93}.
If  large enough, amplification of these perturbations might destabilize the flow and trigger unsteadiness or spatial symmetry breaking, thus leading to subcritical bypass transition.

While eigenvalues are not relevant to characterize such flows, non-modal mechanisms are well described by two complementary ideas: 
transient growth of initial conditions, 
and asymptotic amplification of forcing.
These mechanisms are a result of the non-normality of the linearized Navier-Stokes operator which governs the dynamics of perturbations.
For example, non-normality leads to large transient growth 
in parallel flows through the two-dimensional Orr mechanism and, more importantly, the three-dimensional lift-up effect \cite{But93};
in non-parallel flows, large transient growth is observed because of convective non-normality \cite{Cho05}.
For such flows, transient growth is a well-established notion, and most attempts to control convectively unstable flows naturally focus on reducing the largest possible transient growth, or ``optimal growth''\cite{Cor01tcfd}.
A great variety of control types and strategies exist (see \eg  reviews by Fiedler and Fernholz \cite{Fie90},
Gad-el-Hak \cite{Gad96}, and 
Choi, Jeon and Kim \cite{Choi08}).
Several techniques have been used to reduce transient growth, both active and passive.
Among active control, the design of closed-loop  schemes has received a lot of attention. 
Based on modern control theory (review by Kim and Bewley\cite{Kim07}), 
and  applied to physics-based reduced-order models \cite{Row05, Ehr11} or
to models obtained from system identification \cite{Tia06,Hen07},  
it has proven robust enough to be implemented in experiments. 
Based on Lagrangian optimization, receding-horizon predictive control was able to successfully relaminarize a plane channel flow at a centerline Reynolds number of $1712$ (Bewley, Moin and Temam \cite{Bew01}).
Open-loop control has also been proposed as a successful strategy to mitigate instabilities experimentally, 
be it active or passive (\eg Fransson, Brandt, Talamelli and Cossu \cite{Fra04}, and Pujals, Depardon and Cossu \cite{Puj10b}
to mention a few recent achievements).

As a complementary notion to transient growth,
optimal response to harmonic forcing (or ``optimal gain'') has drawn increasing attention too in the past years.
\r{A}kervik, Ehrenstein, Gallaire and Henningson \cite{Ake08} computed the optimal gain in a flat-plate boundary layer using a reduced-order model of global eigenmodes.
Alizard, Cherubini and Robinet \cite{ali09} used the same method to obtain the optimal gain in a flat plate boundary layer with adverse-pressure-induced separation.
The optimal gain can also be calculated directly from the linearized Navier-Stokes operator, as did 
Garnaud, Lesshafft, Schmid and Huerre \cite{Gar13} for an incompressible axsymmetric jet,
Sipp and Marquet \cite{Sip13} for a flat plate,
and Dergham, Sipp and Robinet \cite{Der13} for a rounded backward-facing step.

Recently, Brandt, Sipp, Pralits and Marquet \cite{Bra11} introduced Lagrangian-based sensitivity analysis to quantify the sensitivity of the largest asymptotic amplification to steady control, and applied it to a flat plate boundary layer. 
Lagrangian-based sensitivity analysis is a variational formulation which allows to compute gradients at low cost through the use of adjoint variables. In particular, it can be applied to flow control with the aim of modifying eigenvalues 
(see Luchini and Bottaro \cite{Luc14} for a recent, general review of adjoint equations).
Hill\cite{Hill92AIAA} derived the corresponding variational formulation and computed the sensitivity of the most unstable eigenvalue to passive control in the incompressible flow past a circular cylinder. He reproduced most sensitive regions previously identified experimentally by Strykowski and Sreenivasan \cite{Stry90}, where a secondary, small control cylinder could suppress vortex shedding.
Sensitivity analysis has then been used by several authors to compute the sensitivity of eigenvalues in absolutely unstable flows.
Marquet, Sipp and Jacquin \cite{mar08cyl} studied the effect of base flow modification and steady control in the bulk in the same flow as Hill\cite{Hill92AIAA} and reproduced most sensitive regions.
Meliga, Sipp and Chomaz \cite{Mel10} managed to control the first oscillating eigenmode  in the compressible flow past an axisymmetric body, considering its sensitivity to steady forcing in the bulk (with mass, momentum or energy sources) and at the wall (with blowing/suction or heating).
Meliga, Pujals and Serre \cite{Mel12} also computed the sensitivity of the shedding frequency (eigenfrequency of the leading global mode to the mean flow) in the fully turbulent wake past a bluff body and reproduced experimental data for the flow forced by a small control cylinder.
The extension of sensitivity analysis to optimal gain by Brandt \etal\cite{Bra11} now opens the way to the control of convectively unstable flows.

In this study, the flow past a wall-mounted bump is considered. This separated flow is characterized by a long recirculation region, high shear, strong backflow, and exhibits large transient growth \cite{Ehr08, Ehr11}. 
Optimal gains are computed at different frequencies, and a sensitivity analysis is systematically performed
 in order to identify regions where they can be reduced with steady open-loop control.
This paper is organized as follows.
Section  \ref{sec:problem} describes the problem, including geometry and governing equations.
The uncontrolled subcritical flow  is studied in section  \ref{sec:noise_amp} by computing linear optimal gains and characterizing  noise amplification with DNS (direct numerical simulation).
In section  \ref{sec:SA}, a sensitivity analysis is performed in order to identify regions where optimal gains are most affected by steady control.
Based on the results, we design one specific control configuration, with wall suction at the bump summit, and demonstrate its effectiveness in reducing not only linear optimal gains  but also non-linear noise amplification.
In section  \ref{sec:stabilization}, we investigate flow stabilization in the supercritical regime: sensitivity analysis applied to global eigenvalues supports the choice of wall suction at the bump summit, which is further confirmed by results from DNS and linear stability analysis.
Conclusions are drawn in section \ref{sec:conclu}.

\section{Problem description and governing equations}
\label{sec:problem}

\begin{figure}
  \centerline{
    \psfrag{y}[][][1][-90]{$y$}
  	\psfrag{x}[][][1][0]{$x$}
  	\psfrag{F}[l][l][1][0]{\small $\FF(t)$}
  	\psfrag{C}[][][1][0]{\small $\CC$}
 	\psfrag{Uc}[][][1][0]{\small $\UU_c$}
 	\psfrag{Ub}[l][l][1][0]{\small $U_{Blasius}$}
  	\includegraphics[width=13 cm]{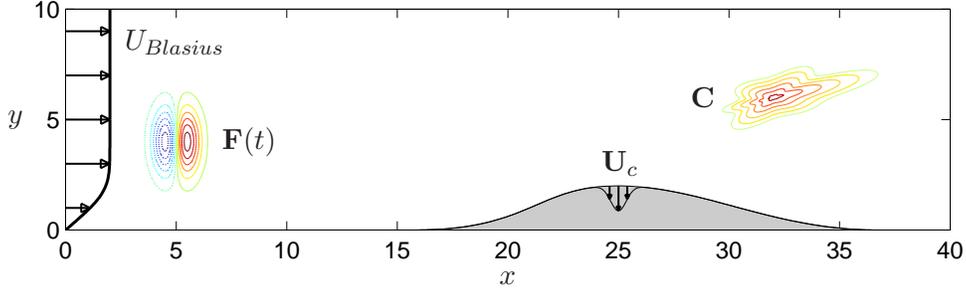}
  }
  \caption{Bump geometry $y=y_b(x)$,
  inlet velocity profile $(U,V)=(U_{Blasius},0)$, 
  time-dependent forcing $\FF(t)$, 
  steady volume control $\CC$
  and steady wall control $\UU_c$.
   }
   \label{fig:bump}
\end{figure}

The flow past a 2D bump mounted on a flat plate is considered. The bump geometry $y=y_b(x)$ is shown in figure \ref{fig:bump} and is the same as in 
Bernard, Foucaut, Dupont and Stanislas \cite{Ber03}, 
Marquillie and Ehrenstein\cite{Mar03} and following studies \cite{Ehr08, Ehr11, pas12}. 
The incoming flow has a Blasius boundary layer profile, characterized by its displacement thickness $\delta^*$ at the reference position $x=0$.
The bump summit is located at $x = x_b =25 \delta^*$, and the bump height is $h=2\delta^*$.
All quantities in the problem are made dimensionless with  inlet velocity $U_\infty$ and inlet boundary layer displacement thickness $\delta^*$.
The Reynolds number is defined as $\Rey=U_\infty \delta^*/\nu$, with $\nu$ the fluid kinematic viscosity.

The fluid motion in the domain $\Omega$ is described by the velocity field $\UU=(U,V)^T$ and  pressure field $P$.
The state vector $\QQ=(\UU,P)^T$ is solution of the two-dimensional incompressible Navier--Stokes equations
\begin{eqnarray}
	& \bnabla \bcdot \UU = 0,
	\quad\quad
	\partial_t \UU +  \bnabla \UU \bcdot  \UU + \nabla  P - \nnu \bnabla^2  \UU =  \FF + \CC \quad \mbox{ in } \Omega, \nonumber\\
	& \UU=\UU_c \quad \mbox{ on } \Gamma_c, 	\label{eq:NS}\\
	& \UU=\00 \quad \mbox{ on } \Gamma_w \setminus \Gamma_c. \nonumber
\end{eqnarray}
In the most general case, $\FF(t)$ is a time-dependent volume forcing, which will be specified as harmonic forcing or stochastic noise in sections \ref{sec:noise_amp} to \ref{sec:stabilization}.
A steady control is applied through a volume force $\CC$ in the flow, or through blowing/suction velocity $\UU_c$  in some regions $\Gamma_c$ of the wall. 
The no-slip condition applies on other parts of the wall $ \Gamma_w \setminus \Gamma_c$.

Without forcing ($\FF=\00$), the steady-state base flow $\QQ_b=(\UU_b,P_b)^T$ satisfies:
\begin{eqnarray}
&	\bnabla \bcdot \UU_b = 0,
	\quad\quad
	 \bnabla \UU_b \bcdot  \UU_b + \nabla  P_b - \nnu \bnabla^2  \UU_b =  \CC \quad \mbox{ in } \Omega, \nonumber\\
	& \UU_b=\UU_c \quad \mbox{ on } \Gamma_c, 	\label{eq:BF}\\
	& \UU_b=\00 \quad \mbox{ on } \Gamma_w \setminus \Gamma_c. \nonumber
\end{eqnarray}

To obtain the equation which govern the evolution of perturbations 
under small-amplitude forcing $\FF=\ff'$,
the flow is written as the superposition of the steady-state base flow and small time-dependent perturbations, 
$\QQ = \QQ_b + \qq'$.
Linearizing 
equations~(\ref{eq:NS}) yields:
\begin{eqnarray}
&	\bnabla \bcdot \uu' = 0, 
	\quad\quad
	\partial_t \uu' + \bnabla \uu' \bcdot \UU_b + \bnabla \UU_b \bcdot \uu' + \bnabla p' - \nnu \bnabla^2 \uu' = \ff' \quad \mbox{ in } \Omega, \nonumber\\
	& \uu'=\00 \quad \mbox{ on } \Gamma_w.
 \label{eq:LN}
\end{eqnarray}

\section{Response to forcing: noise amplification}
\label{sec:noise_amp}

\subsection{Base flow}

The steady-state base flow $\QQ_b$  is obtained with an iterative Newton method. 
A two-dimensional  triangulation of the computational domain $\Omega$ 
($0 \leq x \leq 400$, $y_b \leq y \leq 50$) 
is generated with the finite element software \textit{FreeFem++} (http://www.freefem.org),
and equations (\ref{eq:BF}) are solved in their variational formulation, with the 
following boundary conditions:
Blasius profile $\UU_b=(U_{Blasius},0)^T$ at the inlet,
no-slip condition $\UU_b=\00$  on the wall,
symmetry condition $\partial_y U_b=V_b=0$ at the top border,
and $-P_b\nn+\nnu\bnabla\UU_b\bcdot\nn=\00$ at the outlet, with $\nn$ the  outward unit normal vector.
P2 and P1 Taylor-Hood elements are used for spatial discretization of  velocity and pressure, respectively.

\begin{figure}
  \centerline{
    \psfrag{lc}[r][][1][-90]{$\rl$}
  	\psfrag{re}[t][][1][0]{$\Rey$}
\includegraphics[height=6 cm]{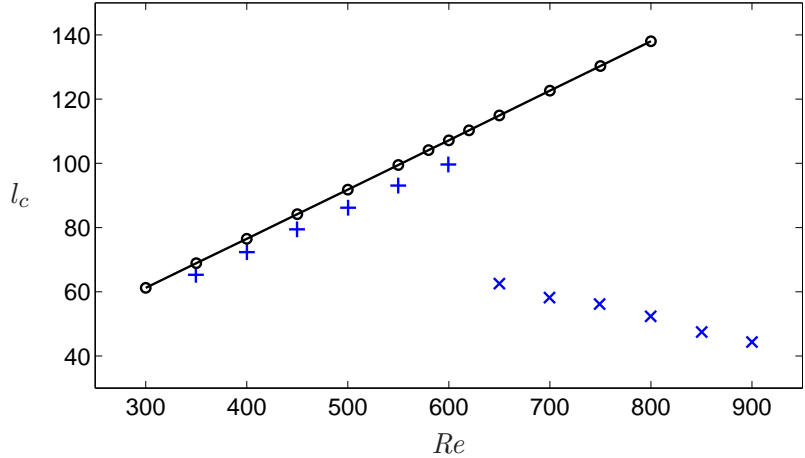}
  }
  \caption{Recirculation length as function of Reynolds number.
  Solid line: steady-state base flow calculated in the present study.
  Symbols: steady state computations ($+$) and
  time-averaged recirculation length of oscillatory flow field ($\times$) obtained by    Marquillie and Ehrenstein \cite{Mar03}.
   }
   \label{fig:lc}
\end{figure}

Figure \ref{fig:lc} shows the recirculation length obtained for different Reynolds numbers.
It can be seen that $\rl$ increases linearly with $\Rey$, a behavior already observed experimentally and numerically in a variety of separated flows, both wall-bounded  and  past bluff bodies: 
backward-facing step\cite{Sin81}, 
bump\cite{pas12},
cylinder\cite{Zie97, Gia07}, etc.
The value of $\rl$ at $\Rey=500$ and $600$ changed by 0.15\% and 0.10\% when refining the computational mesh so as to increase the number of elements by 50\%.
Results from  DNS by Marquillie and Ehrenstein\cite{Mar03} are also reported for reference, where  values up to $\Rey \leq 600$ correspond to steady state computations  and those for $\Rey > 600$ are obtained from time-averaged oscillatory flow fields.
Slight differences stem from different choices of domain size and boundary conditions: in their direct numerical simulations, 
 outlet and upper boundaries 
are located at $x=200$ and $y=80$, 
and the boundary conditions  are respectively $(U,V)=(1,0)$ and an outflow advection condition well suited for DNS.
In the present study, the upper boundary is lower ($y=50$) and the outlet much farther ($x=400$), 
and a stress-free boundary condition is prescribed at both boundaries since it is adapted to steady-state flows and appears as a natural condition when using finite elements.
The present Newton method allows to obtain base flows well beyond the critical Reynolds number and to confirm the linear dependency of $\rl$ with $\Rey$.
%

\subsection{Optimal gain}
\label{sec:optgain}

When harmonic forcing 
$\FF(x,y,t)=\ff(x,y) e^{i\omega t}$
is applied to a stable flow, the asymptotic response is harmonic at the same frequency $\omega$,  $\qq'(x,y,t)=\qq(x,y) e^{i\omega t}$.
Then (\ref{eq:LN}) becomes:
\be
	\bnabla \bcdot \uu = 0, 
	\quad\quad
	i \omega \uu + \bnabla \uu \bcdot \UU_b + \bnabla \UU_b \bcdot \uu + \bnabla p - \nnu \bnabla^2 \uu = \ff.
 \label{eq:HARMFORC}
\ee
In the following, the amplitude of perturbations $\qq$ is measured in terms of their kinetic energy 
$E_p = \int_{\Omega} |\uu|^2 \,\mathrm{d}\Omega = ||\uu||_2^2$ 
with $||.||_2 $ the $L^2$ norm  induced by the Hermitian inner product 
$(\aa|\bb) = \int_{\Omega} \aa^* \bcdot \bb \,\mathrm{d}\Omega$.
The forcing amplitude is measured in a similar way with the $L^2$ norm  
$||\ff||_2^2 = \int_{\Omega} |\ff|^2 \,\mathrm{d}\Omega$.
For a given frequency $\omega$ and a given forcing $\ff$, the asymptotic energy amplification is the gain $G(\omega)=||\uu||_2/||\ff||_2$.
In particular, it is of interest to determine 
 the optimal forcing $\ff_{opt}$ which leads to the largest energy amplification, i.e. the optimal gain:
\be
G_{opt}(\omega) = \max_{\ff} \frac{||\uu||_2}{||\ff||_2}.
\label{eq:optgain1}
\ee
In this study, optimal gains are computed using the same procedure as Garnaud \etal\cite{Gar13}
After spatial discretization, the linear dynamical system  (\ref{eq:HARMFORC}) is written as $(i \omega \BB + \LL) \qq = \BB \PP \ff$, with $\PP$ a 
prolongation operator
 from the velocity-only space to the velocity-pressure space.
The optimal gain (\ref{eq:optgain1}) is recast  as
\be
G_{opt}(\omega) = \max_{\ff} \frac{||\qq||_q}{||\ff||_f},
\label{eq:optgain2}
\ee
where the pseudonorm $||\qq||_q^2 = \qq^H \QQ_q  \qq$ 
and the norm $||\ff||_f^2 = \ff^H \QQ_f  \ff$ 
are discretized versions of the $L^2$ norm.
Rearranging the expression of 
$\displaystyle G^2_{opt}(\omega) = 
 \max_{\ff} ||(i \omega \BB + \LL)^{-1} \BB \PP \ff||_q^2 / ||\ff||_f^2$,
the optimal gain  can be expressed as the leading eigenvalue of the Hermitian eigenvalue problem
\be
\QQ_f^{-1} \PP^H \BB^H (i \omega \BB + \LL)^{-H} \QQ_q
(i \omega \BB + \LL)^{-1} \BB \PP \ff = \lambda \ff.
\label{eq:optgain3}
\ee
The operator $(i \omega \BB + \LL)^{-1}$ is sometimes called ``resolvent'' and the optimal gain $G_{opt}(\omega)$ the ``resolvent norm''.
The largest eigenvalue $ \max(\lambda) = G^2_{opt}(\omega)$ and the associated eigenvector $\ff_{opt}$ are computed using an  implicitly restarted Arnoldi method.
Operators $\QQ_f^{-1}$  and $(i \omega \BB + \LL)^{-1}$  are obtained via  LU decompositions.
Operators $\LL$ and $\BB$  are obtained 
by spatially discretizing the linear system (\ref{eq:HARMFORC}) 
with the same method, same mesh  and same elements as for the base flow.


\begin{figure}[h]
  \centerline{
   	\psfrag{optimal gain}[r][][1][-90]{$G_{opt}$}  	
  	\psfrag{Gmax}[][][1][-90]{$G_{max} \quad\,\,$}
  	\psfrag{ommax}[][][1][-90]{$\omega_{max} \quad$}
  	\psfrag{omega}[t][][1][0]{$\omega$}
  	\psfrag{re}[t][][1][0]{$\Rey$}
  	\psfrag{Re=300}[][][1][0]{\footnotesize $\Rey=300$}
  	\psfrag{Re=580}[][][1][0]{\footnotesize $\Rey=580$}
  	\begin{overpic}[width=15 cm, tics=10]{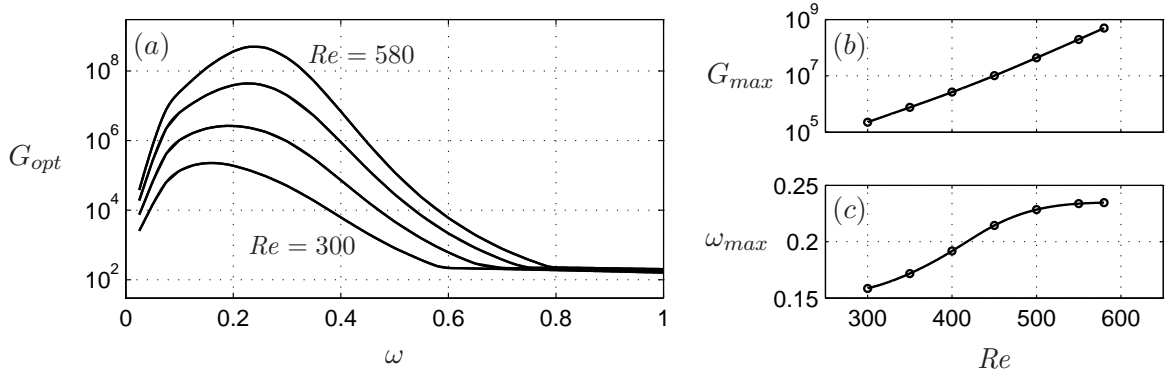}
  	\put(8,27.5){$(a)$}
  	\put(70,27.5){$(b)$}
  	\put(70,13){$(c)$}
  	\end{overpic}
  }
  \caption{
$(a)$ Optimal linear gain at $\Rey=300$, 400, 500 and 580.
$(b)$ Variation of the maximal optimal gain with Reynolds number, and 
$(c)$ frequency of  this maximum.
   }
   \label{fig:optgain}
\end{figure}

Previous studies using DNS \cite{Mar03} and linear global stability analysis \cite{Ehr08} reported a critical Reynolds number $\Rey_c$ between
590 and 610.
In this section, noise amplification is calculated for the subcritical bump flow at Reynolds numbers $\Rey \leq 580$. 

Figure \ref{fig:optgain} shows the optimal gain 
$G_{opt}(\omega)$,  
 its  maximum value $\displaystyle G_{max} = \max_{\omega}(G_{opt}(\omega))$
and the corresponding frequency $\omega_{max}$.
The latter increases between 0.15 and 0.25, while the
maximal optimal gain increases exponentially  with $\Rey$ and reaches values larger than $10^8$ at $\Rey=580$.
This is in agreement with observations for other separated flows, for example pressure-induced recirculation bubbles \cite{ali09}.
The large gain values found here suggest that an incoming noise of very low amplitude  might be linearly amplified so much that it would eventually become of order one and possibly  modify the base flow, or even completely destabilize the overall flow behavior.

Largest values of optimal gain are obtained for frequencies corresponding to the most unstable global eigenvalues (\eg $ 0.15 \leq \omega \leq 0.30$ for the eigenvalues with largest growth rate near critical conditions in Ehrenstein and Gallaire \cite{Ehr08}).
This is also true at lower values of $\Rey$ even though these eigenvalues are strongly stable, a phenomenon known as ``pseudoresonance'' and a direct consequence of non-normality \cite{Tre93}.
No peak is found at the low frequency corresponding to the flapping observed in DNS ($\omega \simeq 0.04$ in Marquillie and Ehrenstein\cite{Mar03}).

\begin{figure}[]
  \centerline{
  	\begin{overpic}[height=6.2 cm, tics=10]{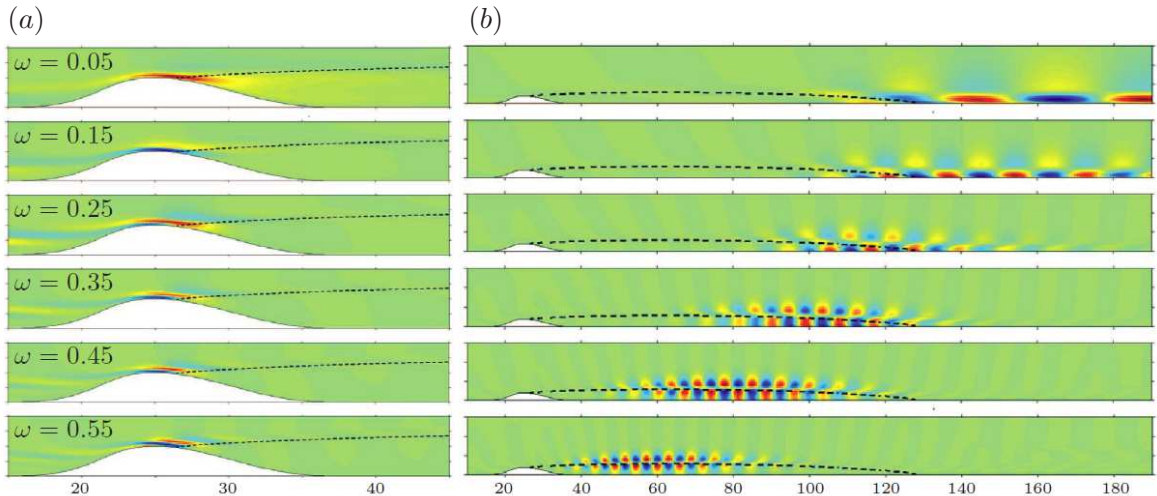}    	
  		\put(1, 41)  {$(a)$}
  		\put(40.5, 41)  {$(b)$}
  	\end{overpic}
  }
  \caption{
  $(a)$ Optimal forcing  and $(b)$ optimal response at $\Rey=580$ for different frequencies $\omega$. The real part of  the streamwise component is shown. The dashed line shows  the base flow separating streamline.
   }
   \label{fig:optforc-optresp}
\end{figure}

Figure \ref{fig:optforc-optresp} shows the spatial structure of the optimal forcing and optimal response at $\Rey=580$.
The optimal forcing is  located near the summit of the bump and at the beginning of the shear layer, with structures tilted against the base flow
(which points to a contribution from the Orr mechanism to the total amplification \cite{Ake08, Der13}).
The forcing exhibits a layer-like structure in the $y$ direction, and these layers become thinner as $\omega$ increases.
The optimal response has a wave packet-like structure in the $x$ direction, whose wavelength decreases with $\omega$.
At low frequency, $\omega \lesssim 0.1$, the response is located 
downstream of the reattachment point and is typical of the convective Tollmien-Schlichting instability \cite{Ehr05, Ake08, Ali07, Der13}.
At intermediate frequencies, the response is located along the shear layer, and its structure is reminiscent of the most unstable global eigenmodes for the same flow\cite{Ehr08}, typical of Kelvin-Helmholtz instability in shear flows.
This intermediate range includes frequencies of largest optimal gain $G_{opt}$ (see figure \ref{fig:optgain}).
At higher frequency, $\omega \gtrsim 0.8$, the optimal forcing and response (not shown) are spread over a wide region and correspond to the combined effect of advection and diffusion \cite{Der13}.

\subsection{Direct numerical simulations}
\label{sec:DNS}

\begin{figure}[h]
  \centerline{
   	\psfrag{y}[][][1][-90]{$y$}
  	\psfrag{x}[t][][1][0]{$x$}
  	\begin{overpic}[height=5.5 cm, tics=10]{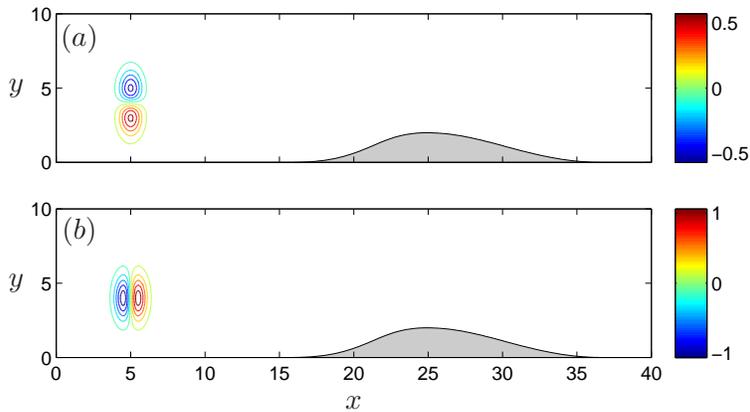}
  	  \put(7,48) { $(a)$}
  	  \put(8.5,22.5) {$(b)$}
  	\end{overpic}
  }
  \caption{
Spatial structure of the divergence-free Gaussian forcing (\ref{eq:actuator}): 
$(a)$ streamwise  and $(b)$ cross-stream  components.
   }
   \label{fig:actuator}
\end{figure}

In this section, the full non-linear Navier-Stokes system  (\ref{eq:NS}) is solved
with direct numerical simulations, using the same procedure as Marquillie and Ehrenstein\cite{Mar03}.
In the following, the subcritical flow at $\Rey=580$ is forced with $\FF = \ff(x,y) \phi(t)$.
This volume forcing will serve a twofold role:
section \ref{sec:DNS_harmonic} focuses on harmonic forcing, so as to investigate the fully non-linear asymptotic response, 
while
section \ref{sec:DNS_stochastic} deals with stochastic forcing, in order to mimic  random noise.
The spatial structure of the forcing is chosen as a divergence-free ``double Gaussian'' already used by Ehrenstein \etal\cite{Ehr11}
and illustrated in figure~\ref{fig:actuator}:
\begin{eqnarray}
& \displaystyle{
f_x  = \,\,\,\, - (y-y_f) A \exp \left( - \frac{(x-x_f)^2}{2\sigma_x^2} -\frac{(y-y_f)^2}{2\sigma_y^2} \right) },
\nonumber \\
& \displaystyle{
f_y =  \frac{\sigma_y^2}{\sigma_x^2} (x-x_f) A \exp \left( -\frac{(x-x_f)^2}{2\sigma_x^2} -\frac{(y-y_f)^2}{2\sigma_y^2} \right). }
\label{eq:actuator}
\end{eqnarray}
with a variable amplitude $A$, a center located at  $x_f=5$, $y_f=4$, and characteristic width and height
$\sigma_x=0.5$, $\sigma_y=1.0$.
The Gaussian-type forcing $\ff$ is sufficiently far from the wall so that its 
 $L^2$ norm is very close to the theoretical value
$  A \sqrt{ \frac{\pi}{2} \sigma_x \sigma_y^3\left( 1+\sigma_y^2/\sigma_x^2  \right)  }$  one would obtain in an unbounded domain, yielding $||\ff|| \simeq 2 A$.
It should be stressed that $\FF(t)$ aims at modelling an external forcing,
and should not be confused with volume control $\CC$ or wall control $\UU_c$.

\subsubsection{DNS with harmonic forcing}
\label{sec:DNS_harmonic}

\begin{figure}[]
  \centerline{
    \psfrag{E}[][][1][-90]{$E_p \,\,$}
   	\psfrag{Ep}[][][1][-90]{$\overline E_p \,\,$}
  	\psfrag{t}[t][][1][0]{$t$}
  	\psfrag{A=1e-4}[bl][bl][1][0]{\footnotesize $A=10^{-4}$}
  	\psfrag{A=1e-5}[bl][bl][1][0]{\footnotesize $A=10^{-5}$}
  	\psfrag{A=1e-6}[bl][bl][1][0]{\footnotesize $A=10^{-6}$}
  	\psfrag{A=1e-7}[bl][bl][1][0]{\footnotesize $A=10^{-7}$}
  	\psfrag{A=1e-8}[bl][bl][1][0]{\footnotesize $A=10^{-8}$}
  	\psfrag{A}[][][1][0]{$A$}
  	\psfrag{s1}[b][b][1][0]{\footnotesize 1}
 	\psfrag{s2}[][][1][0]{\footnotesize \,\,\,2}
    \begin{overpic}[height=6 cm, tics=10]{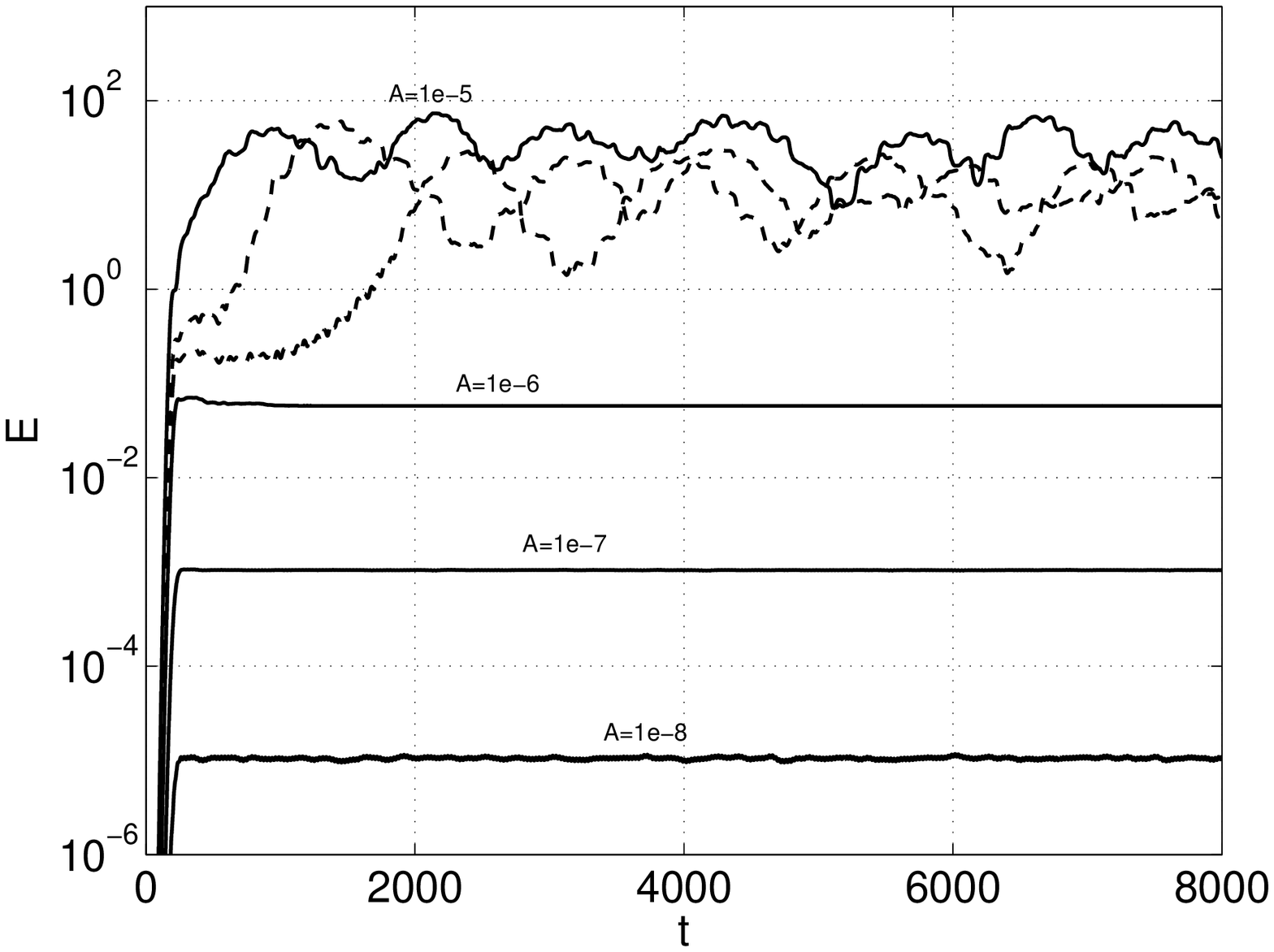}
	  \put(0,78){$(a)$}
	\end{overpic}
  	\begin{overpic}[height=6 cm, tics=10]{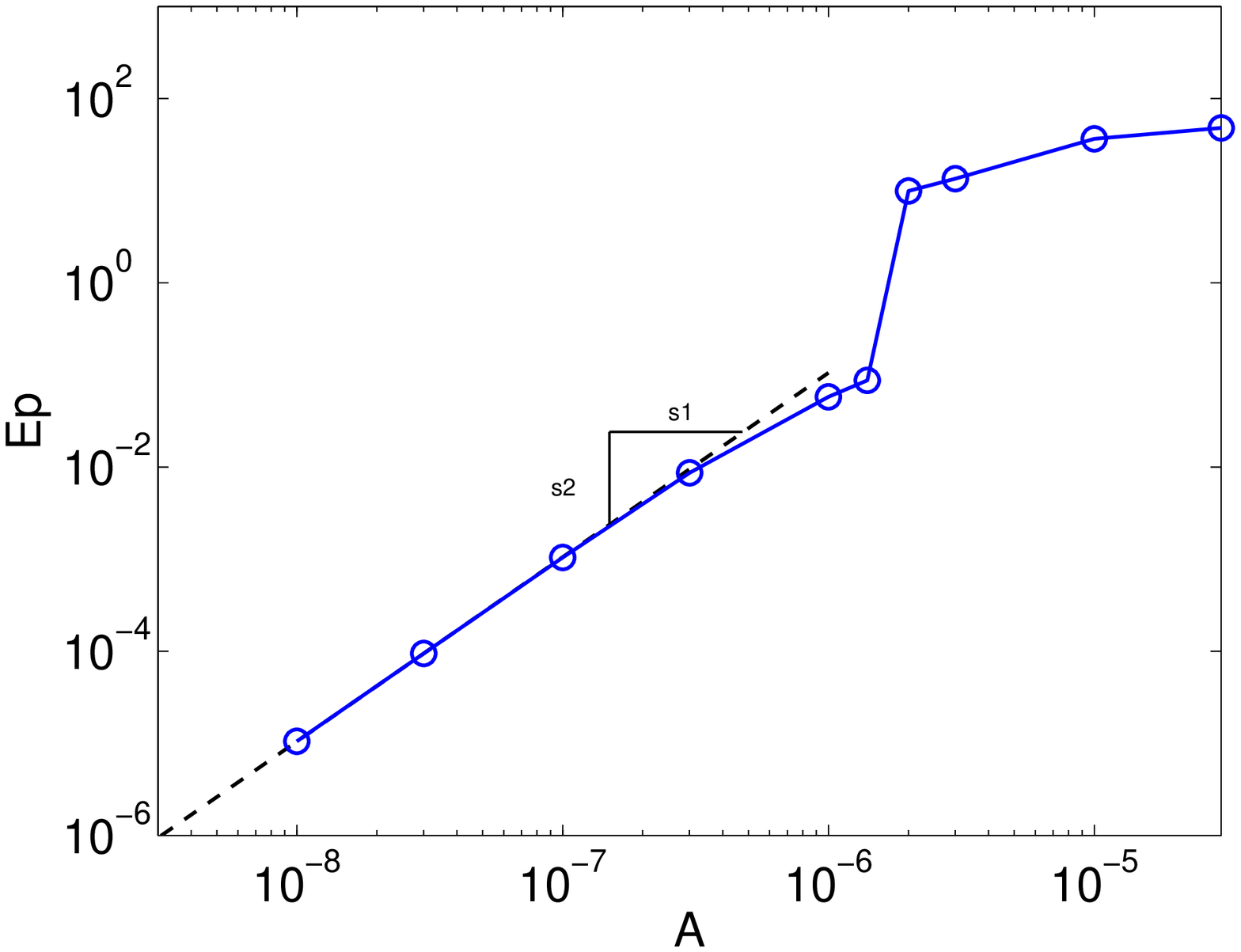}
  	  \put(0,80){$(b)$}
	\end{overpic}
  }
  \vspace{0.9cm}
  \centerline{
  	\psfrag{U4}[][t][1][-90]{$ u'/A $}
  	\psfrag{U3}[][t][1][-90]{$ u'/A $}
  	\psfrag{U2}[][t][1][-90]{$ u'/A $}
  	\psfrag{t}[t][][1][0]{$t$}
    \begin{overpic}[width=11.8 cm, tics=10]{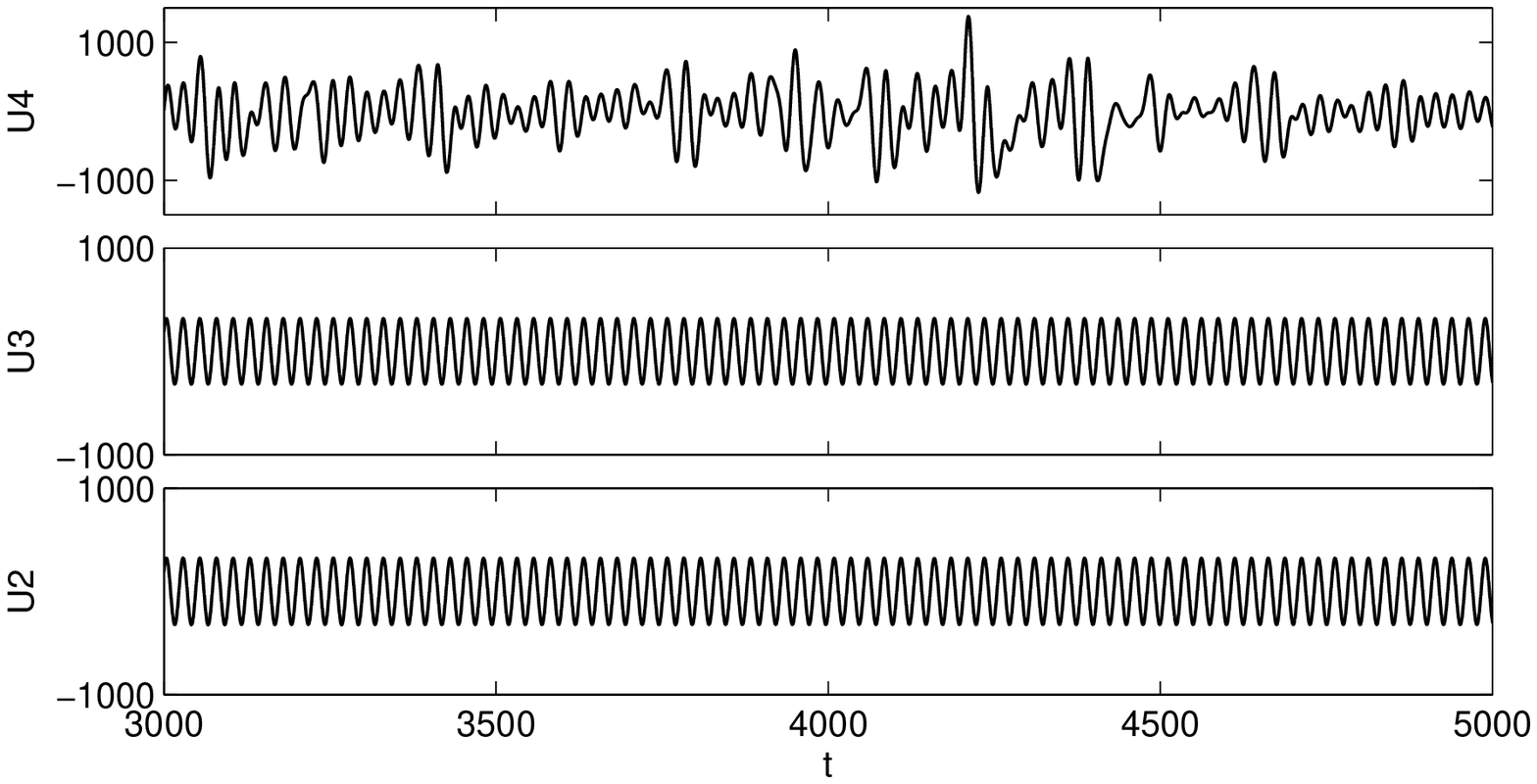}
	\put(12,52){$(c)$}
	\put(15,47){\footnotesize $A=10^{-5}$}
	\put(15,31.5){\footnotesize $A=10^{-6}$}
	\put(15,16){\footnotesize $A=10^{-7}$}
	\end{overpic}
  }
  \vspace{0.8cm}
  \centerline{
    \hspace{0.4 cm}
  	\psfrag{ps}[][t][1][0]{$\log_{10}$PS}
  	\psfrag{om}[t][][1][0]{$\omega$}
    \begin{overpic}[width=10.7 cm, tics=10]{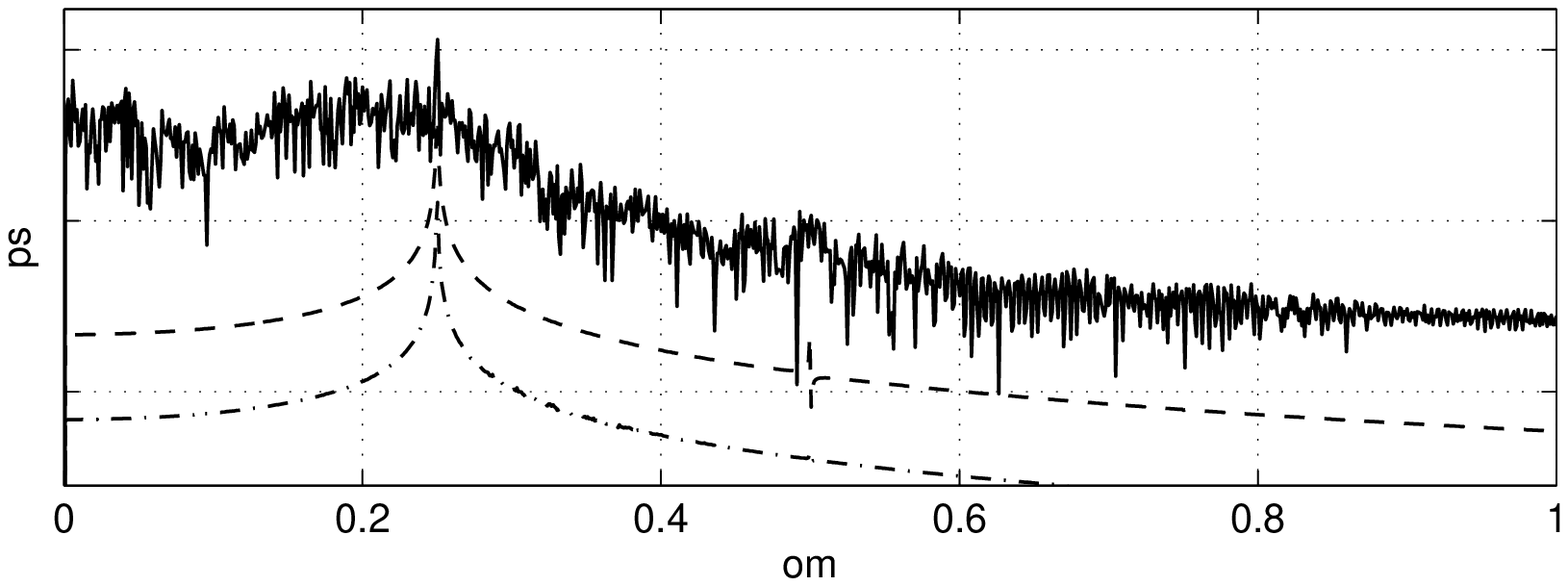}
	\put(5,39.5){$(d)$}
	\put(7,32.5){\footnotesize $A=10^{-5}$}
	\put(7,18)  {\footnotesize $A=10^{-6}$}
	\put(7,12.5){\footnotesize $A=10^{-7}$}
	\end{overpic}
  }
  \caption{
  Response to  harmonic forcing at $\Rey=580$, $\omega=0.25$.
$(a)$ 
Time evolution of the energy of the perturbations.
Dashed lines correspond to $A=2\times10^{-6}$ and $3\times10^{-6}$;
$(b)$ 
Mean asymptotic energy in the steady-state regime as function of the forcing amplitude $A$. The solid line has a slope 2;
 $(c)$ 
 Time series of the streamwise perturbation velocity $u'$ at $x=80, y=1$, for $A=10^{-7}$, $10^{-6}$, $10^{-5}$;
  $(d)$ 
 Power spectrum of this velocity for forcing amplitudes 
 $A=10^{-7}$ (dash-dotted line), 
 $A=10^{-6}$ (dashed line),
 $A=10^{-5}$ (solid line)
(arbitrary unit, logarithmic scale).
   }
 \label{fig:E_harm}
\end{figure}

\begin{figure}
  \centerline{
   	\psfrag{G}[][][1][-90]{$G$}
  	\psfrag{omega}[t][][1][0]{$\omega$}	
  	\psfrag{Gopt}[l][l][1][0]{\footnotesize $G_{opt} $}	
  	\psfrag{Gdns}[][][1][0]{\footnotesize $G_{DNS}$}
  	\psfrag{Glin}[][][1][0]{\footnotesize $G_{lin}$}
  	\includegraphics[height=6 cm]{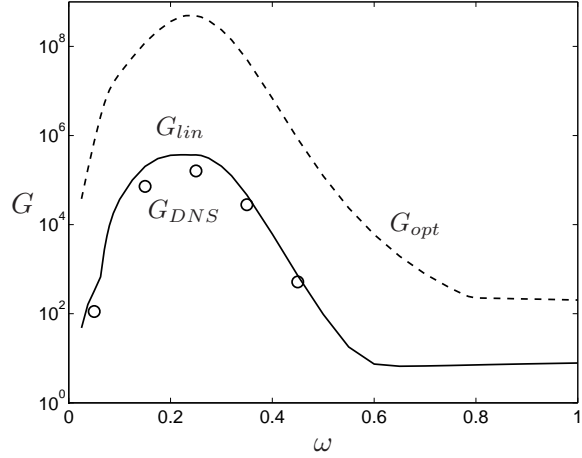}
  }
\caption{
Actual response to harmonic forcing $\FF = \ff(x,y) e^{i \omega t}$ 
with the particular choice (\ref{eq:actuator}) for the spatial structure $\ff$.
Solid line: actual linear gain $G_{lin}$; 
Symbols: linear DNS gain $G_{DNS}$  obtained from DNS calculations with small-amplitude forcing.
The dashed line indicates for reference the  optimal gain $G_{opt}$  (reported from figure \ref{fig:optgain}).
   }
   \label{fig:optgain_DNS}
\end{figure}

In this section the forcing is chosen as $\FF = \ff(x,y) \phi(t)$ with a harmonic time-dependency:
$\phi(t) = e^{i\omega t}$.
We introduce notations for different measures of harmonic amplification used in the following:
\begin{itemize}
\item \textit{Linear optimal gain} (``resolvent norm'')  $G_{opt}(\omega)$, 
already defined by equation (\ref{eq:optgain1}): largest energy amplification over all possible forcings $\ff(x,y)$, it is solution of the eigenvalue problem (\ref{eq:optgain3});
\item \textit{Actual linear gain} $G_{lin}(\omega)$: 
energy amplification actually obtained for our particular choice of forcing (\ref{eq:actuator}) in a fully linearized setting, it is simply calculated by solving the linear system (\ref{eq:HARMFORC}), i.e.
$G_{lin}(\omega)=||\qq||_q/||\ff||_f = 
 ||(i \omega \BB + \LL)^{-1} \BB \PP \ff||_q / ||\ff||_f$;
\item \textit{Linear DNS gain} $G_{DNS}(\omega)$: 
energy amplification $||\qq||/||\ff||$ measured in non-linear DNS forced by our particular choice of forcing  (\ref{eq:actuator}) in the linear regime, i.e. with forcing amplitudes  small enough for non-linear effects to be negligible.
\end{itemize}
Figure \ref{fig:E_harm}$(a)$ displays the evolution of the energy of the perturbations $E_p(t)=||\qq'(t)||^{2}=||\QQ(t)-\QQ_b||^{2}$ for different forcing amplitudes, at $\omega=0.25$. 
For small values of $A$, the flow quickly reaches a steady-state regime, 
as $E_p$ reaches an almost constant value, and 
 the flow is harmonic as indicated by the regular velocity signal and the peaked power spectrum shown in figures \ref{fig:E_harm}$(c,d)$.
From the results in this small-amplitude forcing regime, it is possible to measure the amplification from forcing to response, or \textit{linear DNS gain}  
$G_{DNS} = ||\qq||/||\ff||=E_p^{1/2}/2A$. 
The variation of the mean asymptotic energy $\overline E_p$ (mean value of $E_p(t)$ after the transient regime) with $A$ is shown in figure \ref{fig:E_harm}$(b)$  in logarithmic scale. For small values of $A$, the slope of the curve  is 2 as expected, and the \textit{linear DNS gain} is $G_{DNS} \simeq 1.6 \times 10^5$. This value should be compared to the \textit{actual linear gain}  $G_{lin}(\omega)$.  Values of $G_{lin}$ and $G_{DNS}$ are given in figure \ref{fig:optgain_DNS} and show good agreement.

For larger values of the forcing amplitude, non-linear effects become non-negligible and the energy amplification starts to depart from the linear gain.
At some point, close to $A_c \simeq 2 \times 10^{-6}$ in the case illustrated here, 
transition to a different regime occurs.
The flow is destabilized and becomes non-harmonic, as indicated by figures \ref{fig:E_harm}$(c,d)$: although a sharp peak is still present at the forcing frequency, the field now also contains a whole range of  low and mid frequencies.
The perturbation energy jumps to a larger  value.
This phenomenon is a subcritical transition: 
small finite-amplitude perturbations are large enough to make the initially stable flow move away from a weakly non-linear oscillatory state to a disordered one.
Increasing $A$ further does not modify significantly the mean asymptotic energy, which saturates at $\overline E_p \simeq 100$.

\subsubsection{DNS with stochastic forcing}
\label{sec:DNS_stochastic}

In this section the forcing is chosen as $\FF = \ff(x,y) \phi(t)$ with a stochastic time-dependency:
$\phi(t)$ is a random noise of normal distribution (zero mean, unit variance).
After investigating the response to harmonic forcing and comparing with linear results in 
section \ref{sec:DNS_harmonic}
the aim is now to model a more realistic noise.

\begin{figure}
  \centerline{
  	\hspace{0.1cm}
   	\psfrag{Ep}[][t][1][-90]{$E_p$}
   	\psfrag{Epm}[][][1][-90]{$\overline E_p \,\,$}
  	\psfrag{t}[t][][1][0]{$t$}
  	\psfrag{A}[t][][1][0]{$A$}
  	\psfrag{s1}[][][1][0]{\footnotesize 1}
 	\psfrag{s2}[][][1][0]{\footnotesize 2\,\,\,\,\,\,}
  	\psfrag{A=1e-2}[bl][bl][1][0]{\footnotesize $A=10^{-2}$} 
  	\psfrag{A=1e-3}[bl][bl][1][0]{\footnotesize $A=10^{-3}$}  
  	\psfrag{A=1e-4}[bl][bl][1][0]{\footnotesize $A=10^{-4}$}
  	\psfrag{A=1e-5}[bl][bl][1][0]{\footnotesize $A=10^{-5}$}
  	\psfrag{A=1e-6}[bl][bl][1][0]{\footnotesize $A=10^{-6}$}
  	\psfrag{A=1e-7}[bl][bl][1][0]{\footnotesize $A=10^{-7}$}  
	\begin{overpic}[height=5.8 cm, tics=10]{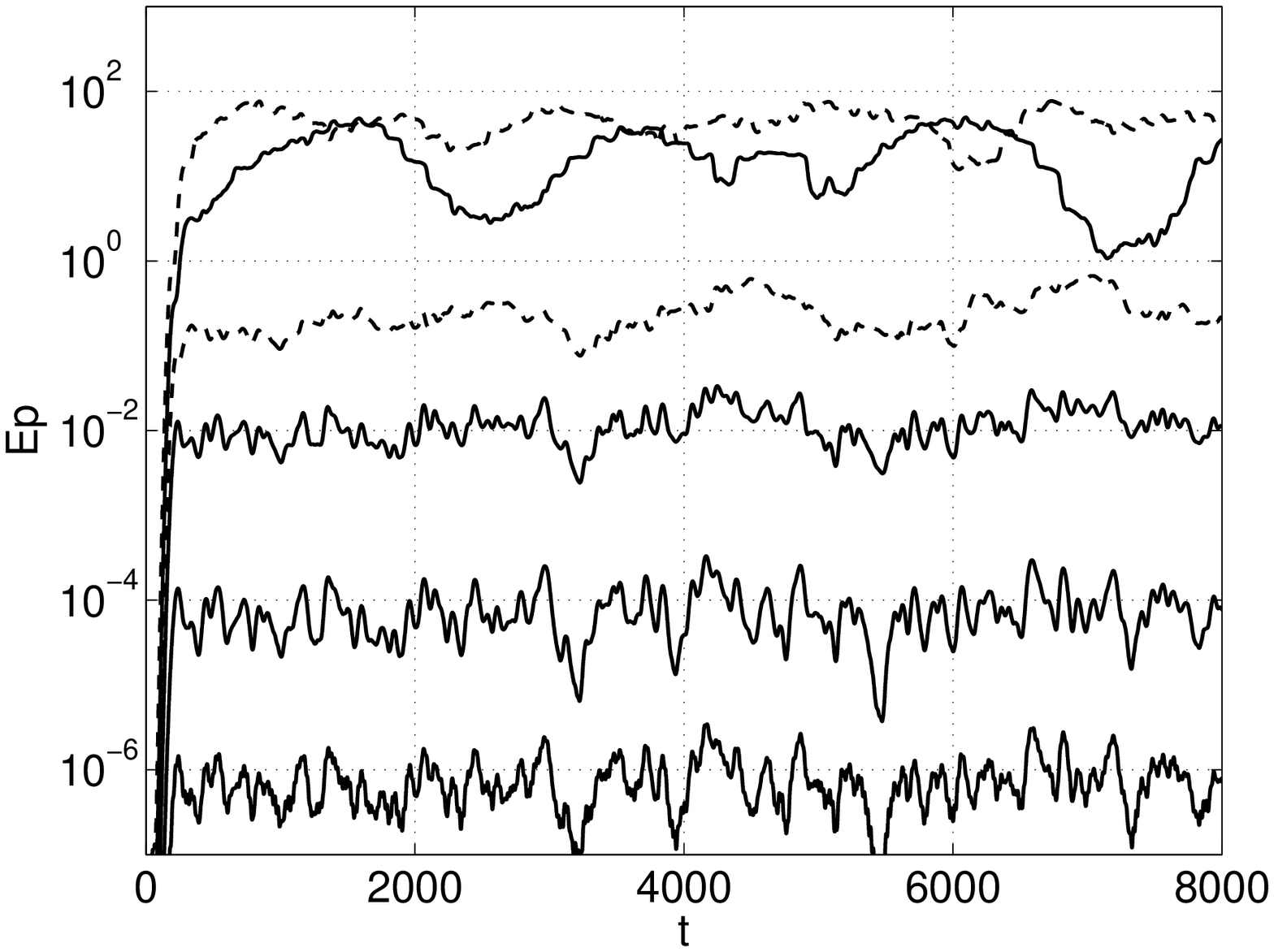}
	\put(12,77){$(a)$}
	\end{overpic}
	\begin{overpic}[height=5.8 cm, tics=10]{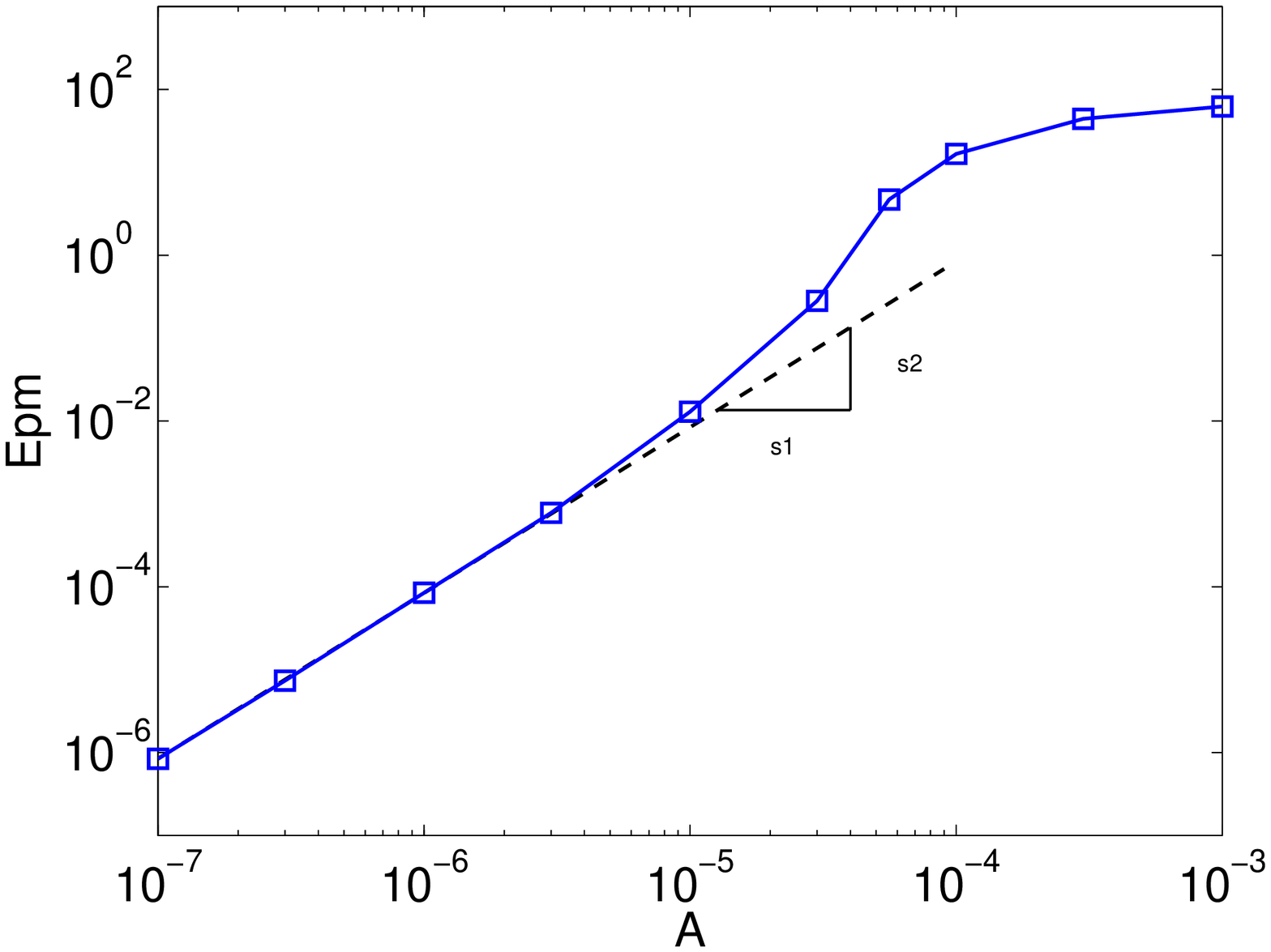}
	\put(13,77){$(b)$}
	\end{overpic}
  }
  \vspace{0.9 cm}
  \centerline{
    \hspace{0.1cm}
  	\psfrag{U80} [][t][1][-90]{$ \displaystyle\frac{u'}{A} $}
  	\psfrag{U140}[][t][1][-90]{$ \,\,\,\,\displaystyle\frac{u'}{A} $}
	\psfrag{t}[t][][1][0]{$t$}
  	\hspace{0.45cm}
    \begin{overpic}[height=4 cm, tics=10]{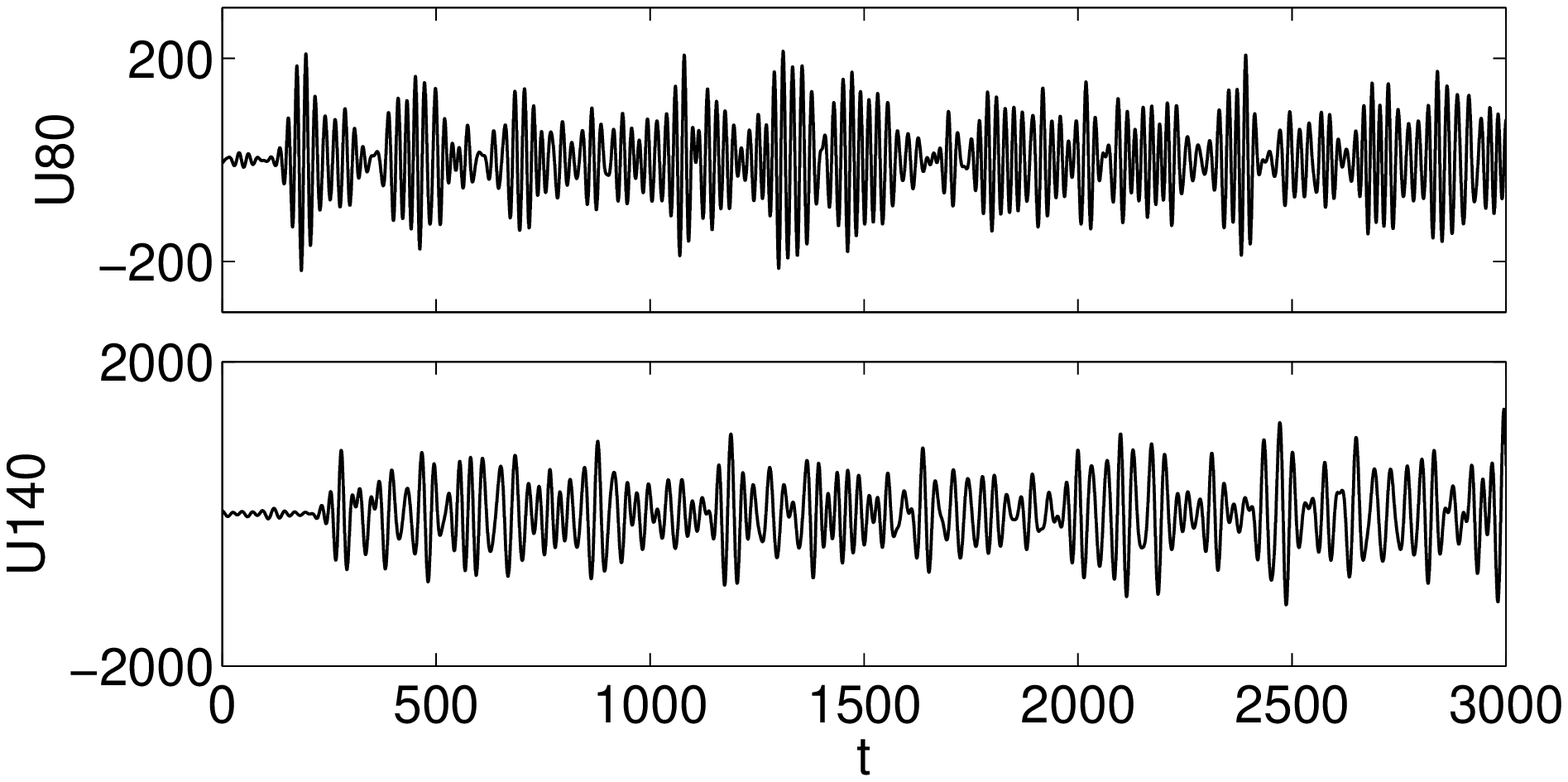}
	\put(14,53){$(c)$}
	\end{overpic}
	\hspace{0.05cm}
    \begin{overpic}[height=4 cm, tics=10]{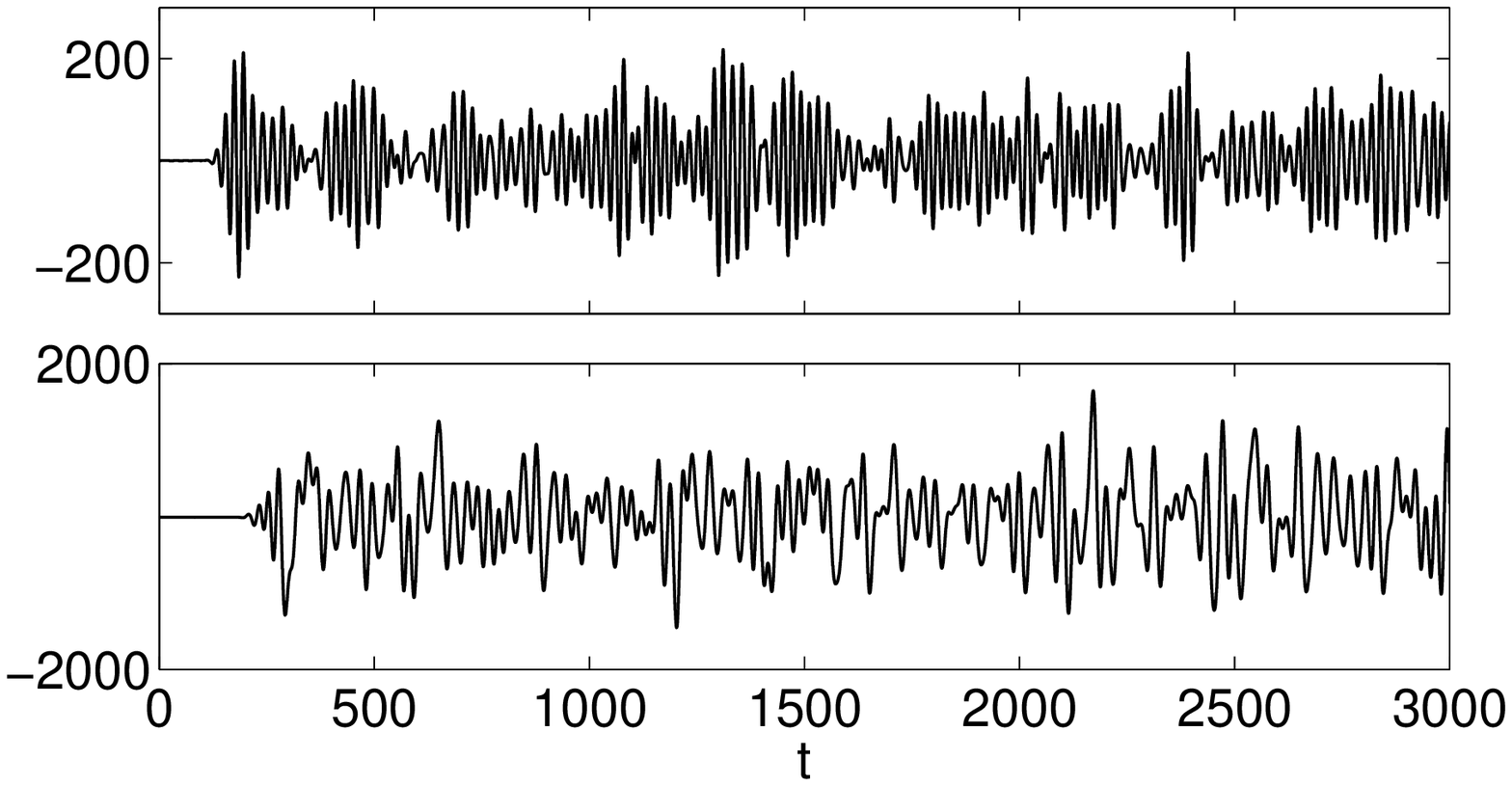}
	\put(12,54){$(d)$}
	\put(78,32.5)	{\footnotesize $x=80$}
	\put(78,9.5)  	{\footnotesize $x=140$}
	\end{overpic}
	\hspace{0.8 cm}
   }
  \vspace{1. cm}
    \centerline{
  	\psfrag{a.u.}[][t][1][0]{\hspace{4cm}PS (a.u.)}
  	\psfrag{om}[t][][1][0]{$\omega$}
  	\hspace{0.25cm}
    \begin{overpic}[height=6 cm, tics=10]{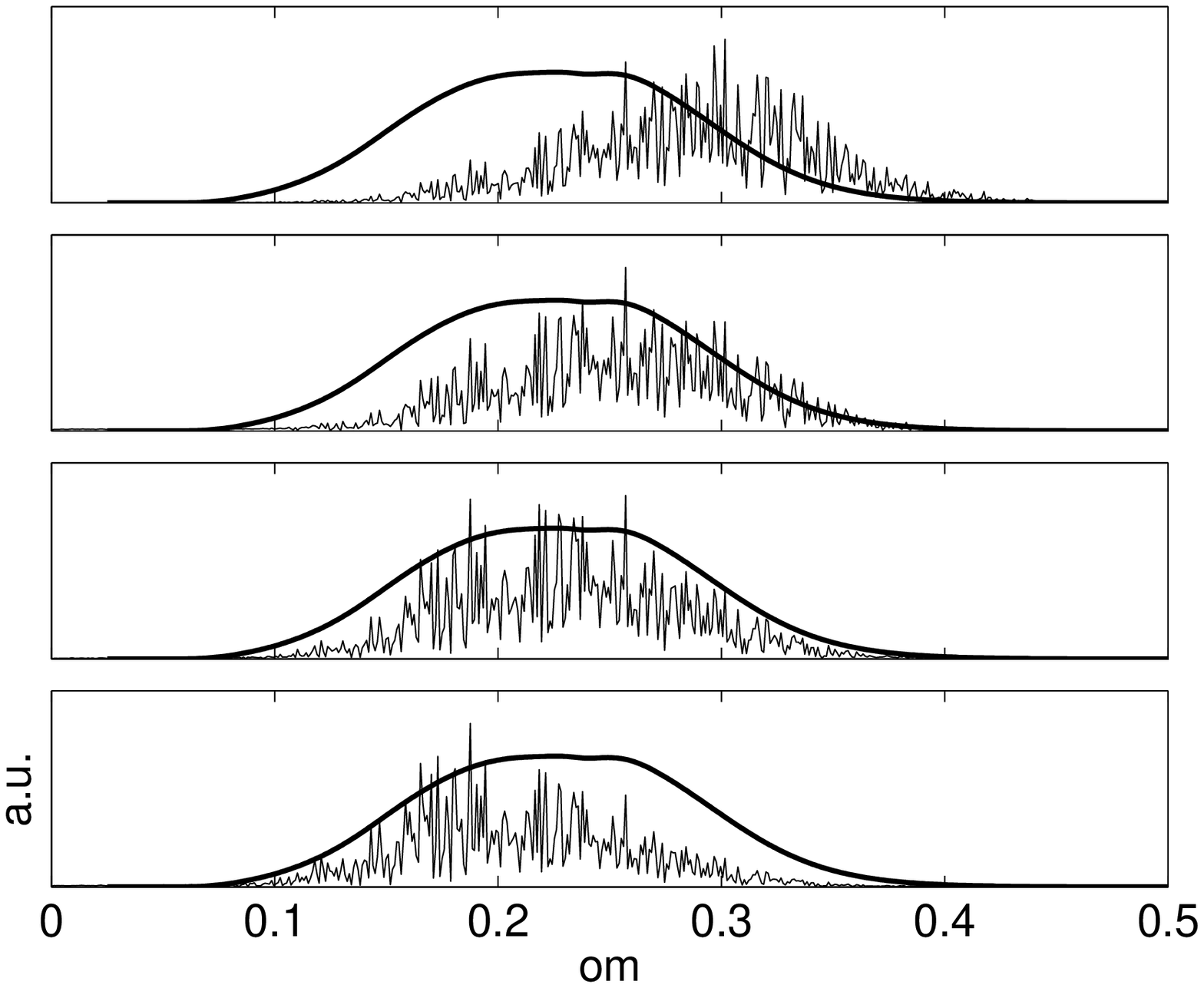}
	\put(6,84){$(e)$}
	\put(78,75){\footnotesize $x=80$}
	\put(78,55.3){\footnotesize $x=100$}
	\put(78,35.6){\footnotesize $x=120$}
	\put(78,16)  {\footnotesize $x=140$}
	\end{overpic}
	\hspace{1cm}
    \begin{overpic}[height=6 cm, tics=10]{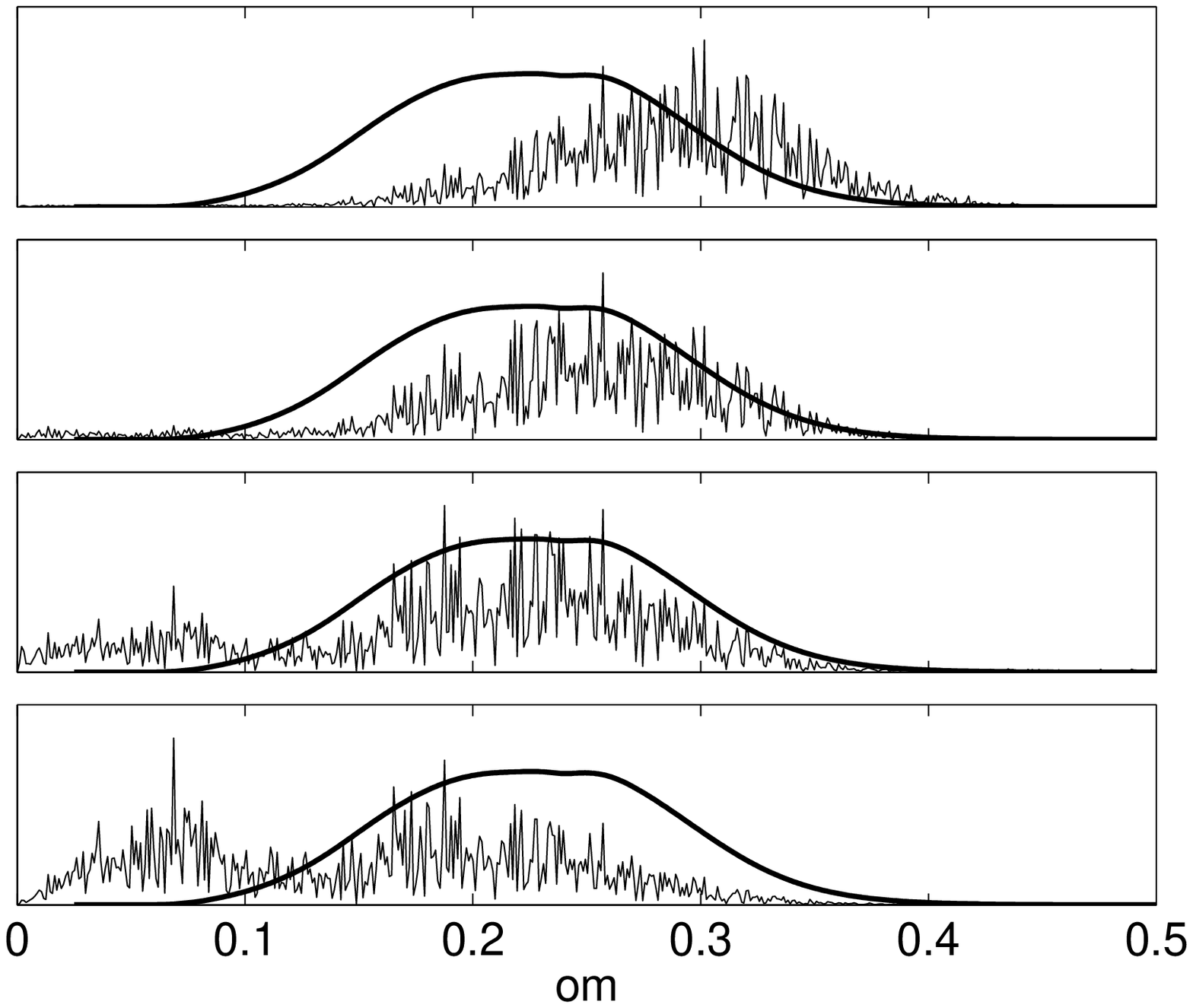}
	\put(2,88){$(f)$}
	\put(78,75){\footnotesize $x=80$}
	\put(78,55.3){\footnotesize $x=100$}
	\put(78,35.6){\footnotesize $x=120$}
	\put(78,16)  {\footnotesize $x=140$}
	\end{overpic}
  }
  \caption{Response to stochastic forcing at $\Rey=580$.
$(a)$ 
Time evolution of the energy of the perturbations  $E_p$.
 Dashed lines correspond to $A=3\times10^{-5}$  and $3\times10^{-4}$;
$(b)$  
 Mean asymptotic energy in the steady-state regime as function of the forcing amplitude $A$;
$(c,d)$ 
 Time series of the streamwise perturbation velocity $u'$ measured at $y=1$ and $x=80$ and 140 for $(c)$ $A=10^{-7}$ and $(d)$ $A=10^{-5}$;
$(e,f)$ 
 Power spectrum of the streamwise velocity measured at $y=1$ and $x=80$, 100, 120, 140, for 
 $(e)$ $A=10^{-7}$ and 
$(f)$  $A=10^{-5}$.
For reference, the thick line shows the (uncontrolled) linear gain  $G_{lin}(\omega)$ from figure \ref{fig:optgain_DNS} (arbitrary unit, linear scale).
   }
   \label{fig:E_rand}
\end{figure}

The time evolution of $E_p(t)$ and the variation of $\overline E_p$ with forcing amplitude are shown in figures \ref{fig:E_rand}$(a,b)$.
Qualitatively, they are very similar to their counterparts for harmonic forcing. 
In particular, $\overline E_p$ is proportional to $A^2$ for small amplitudes, increases more quickly after a critical value $A_c \simeq 10^{-5}$, and  then saturates.
$A_c$ is larger and the transition smoother than in the harmonic forcing case.
This is consistent with the fact that a random white noise excites all frequencies, thus only part of the total forcing energy is available at  amplified frequencies. This results in a larger forcing amplitude needed to obtain the same destabilizing effect.
However, the level of noise that  the system can withstand  is still very low, which shows that the subcritical bump flow  is a strong noise amplifier, easily destabilized by incoming noise \cite{Cho05}.

\begin{figure}
    \centerline{
    \begin{overpic}[width=15.5cm, height=2.7cm, tics=10]{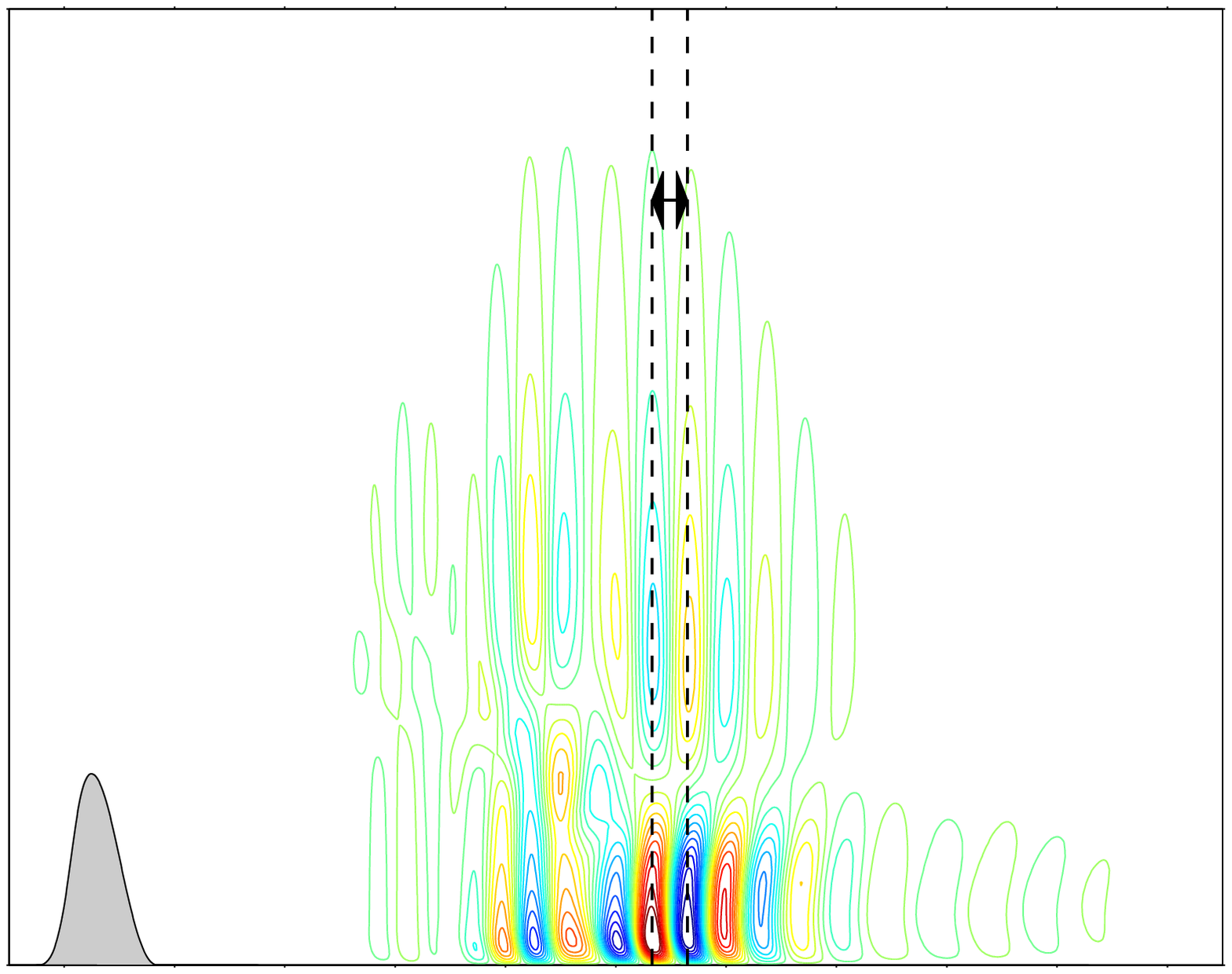}
	\put(2,14){$(a)$ $A=10^{-7}$}
	\put(56.5,13){\footnotesize $\lambda_1/2$}
	\end{overpic}
	}
	\centerline{
    \begin{overpic}[width=15.5cm, height=2.7cm, tics=10]{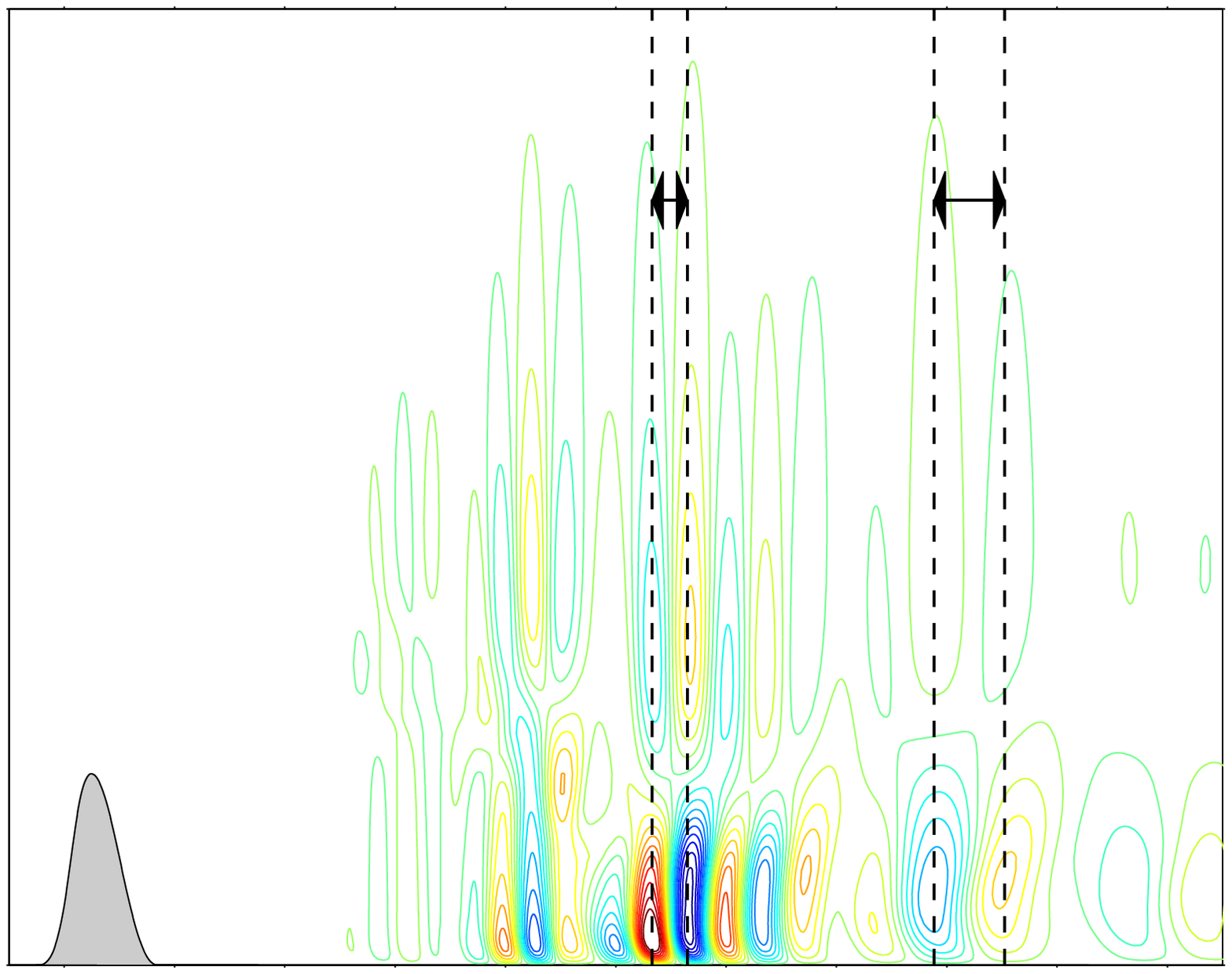}
	\put(2,14){$(b)$ $A=10^{-5}$}
	\put(56.5,13){\footnotesize $\lambda_1/2$}
	\put(82,  13){\footnotesize $\lambda_2/2=\lambda_1$}
	\end{overpic}
	}
	\centerline{
    \begin{overpic}[width=15.5cm, height=2.7cm, tics=10]{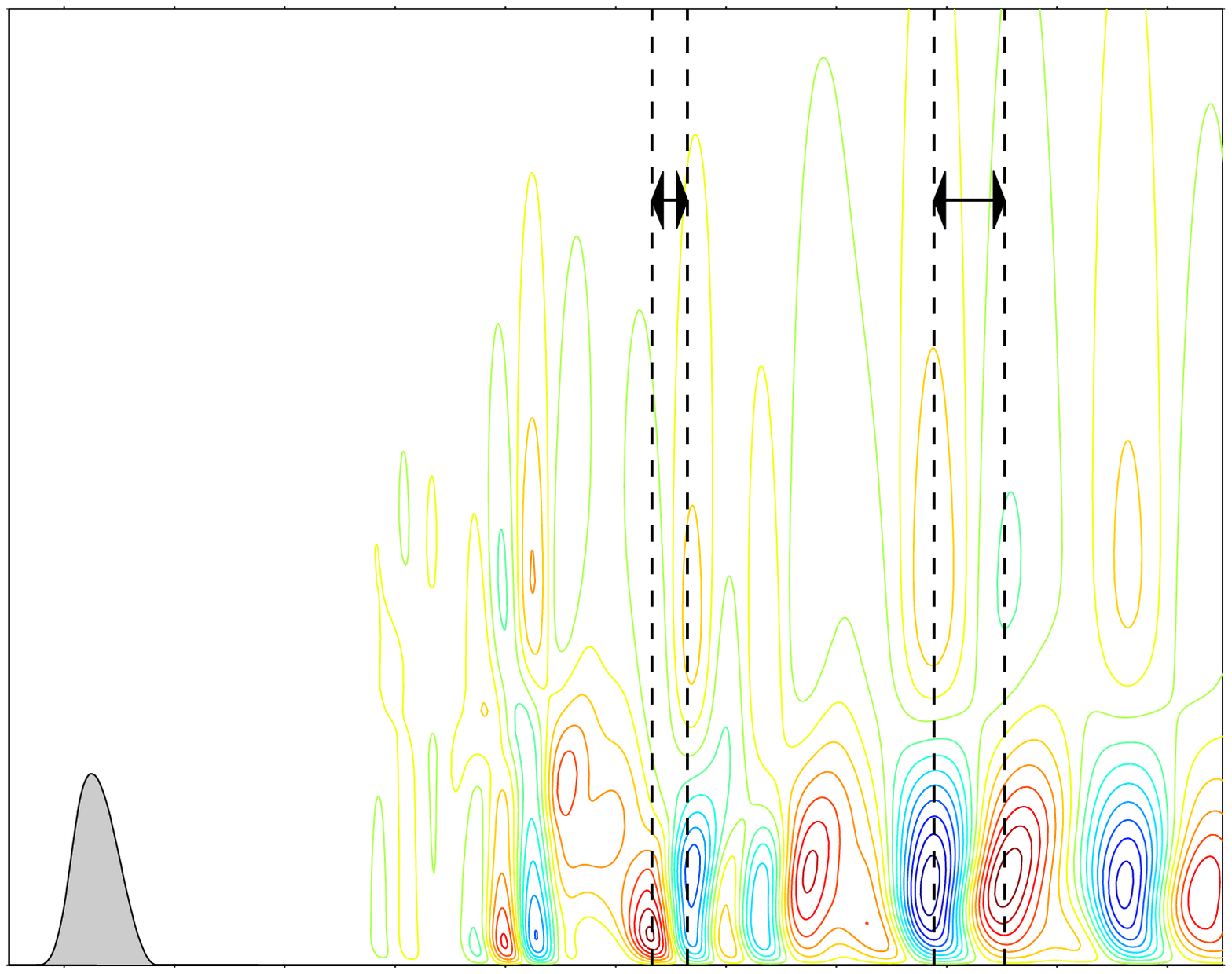}
	\put(2,14){$(c)$ $A=3\times10^{-5}$}	
	\put(56.5,13){\footnotesize $\lambda_1/2$}
	\put(82,  13){\footnotesize $\lambda_2/2=\lambda_1$}
	\end{overpic}
	}
	\centerline{
    \begin{overpic}[width=15.5cm, height=2.7cm, tics=10]{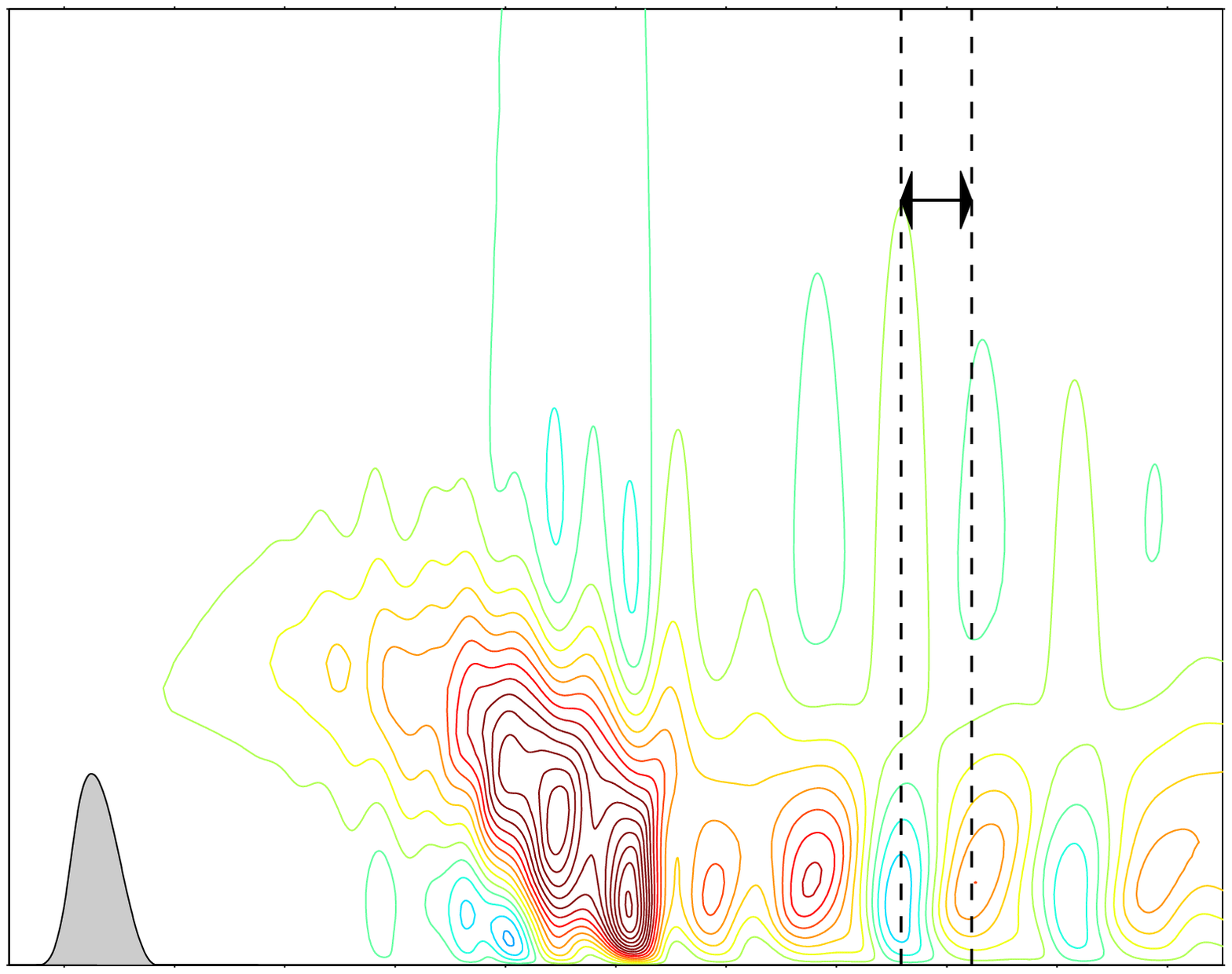}
	\put(2,14){$(d)$ $A=10^{-4}$}	
	\put(79.5,13){\footnotesize $\lambda_2/2=\lambda_1$}
	\end{overpic}
	}
 \caption{
Subharmonic instability occurs as a manifestation of non-linear effects when forcing amplitude is large enough.
Amplitude of the stochastic forcing: 
$(a)$~$A=10^{-7}$, 
$(b)$~$A=10^{-5}$,
$(c)$~$A=3\times10^{-5}$,
$(d)$~$A=10^{-4}$.
Contours of streamwise perturbation  velocity, $t=2000$, $\Rey=580$.
The axes are not to scale.
   }
   \label{fig:vortexpairing}
\end{figure}

Figure \ref{fig:E_rand}$(c$-$f)$ shows streamwise velocity time series and power spectra computed with streamwise velocity signals measured over $1500 \leq t \leq 8000$ at $y=1$, from upstream ($x=80$) to downstream ($x=140$) of the reattachment point.
Interestingly, power spectra shift towards lower frequencies as $x$ increases, which is consistent with the fact that   
linear optimal response moves downstream when $\omega$ decreases (see figure \ref{fig:optforc-optresp}).
For the lower forcing amplitude $A=10^{-7}$, power  spectra in figure \ref{fig:E_rand}$(e)$ are maximal in the range of frequencies $0.15 \lesssim \omega \lesssim 0.30$, close to the range where linear gain (thick curve) is large.
The agreement between linear optimal gain and non-linear DNS power spectra is best at $x=120$, where optimal responses are mostly located for this range of frequencies and in particular at $\omega_{max}=0.23$ (figure \ref{fig:optforc-optresp}).
For the larger forcing amplitude $A=10^{-5}$, 
power spectra in figure \ref{fig:E_rand}$(f)$ 
are the same as those for $A=10^{-7}$ inside the recirculation region.
Downstream, however, they exhibit two distinct groups of frequencies: the same as for $A=10^{-7}$, and another one at lower frequencies, related by a factor 1/2. 
Inspection of velocity fields in this case, shown in figure    \ref{fig:vortexpairing}, reveals 
a secondary subharmonic instability reminiscent of vortex pairing:
structures downstream have a wavelength twice as large as the primary wavelength observed upstream, $\lambda_2=2\lambda_1$.
For larger forcing amplitudes, the flow is more complex because of even stronger non-linear effects, and wavelengths are slightly increased. 
Note that perturbations in the linear regime (figure    \ref{fig:vortexpairing}$(a)$) are very similar to the linear optimal response at  $\omega=0.25$ (figure \ref{fig:optforc-optresp}$(b)$), close to the most amplified frequency.

\section{Sensitivity analysis}
\label{sec:SA}

\subsection{Sensitivity of optimal gain}
\label{sec:sens_opt_gain}

In order to design an efficient open-loop control strategy aiming at reducing the optimal gain, it is important first to understand the effect of a given control on $G_{opt}(\omega)$.
Following Brandt \etal \cite{Bra11}, a variational technique formulated in a Lagrangian framework is used to evaluate the linear sensitivity of the optimal gain with respect to control.
Considering the small variation of $G_{opt}^2(\omega)$ resulting from a small source of momentum 
$\bdelta \CC$ in the domain $\Omega$
and small-amplitude wall blowing/suction $\bdelta \UU_c$ on the control wall $\Gamma_c$,
the sensitivities to these two types of control can be defined as 
$\delta G_{opt}^2 = (\bnabla_\CC G_{opt}^2     | \bdelta \CC)
     + \langle \bnabla_{\UU_c} G_{opt}^2 | \bdelta \UU_c \rangle$,
where the second term 
is  a one-dimensional inner product on the control boundary  $\langle \aa \,|\, \bb \rangle = \int_{\Gamma_c} \aa^* \bcdot \bb \,\mathrm{d}\Gamma$.
Through the use of a Lagrangian that includes the definition of the optimal gain (\ref{eq:optgain2}), one  finds the expressions
\begin{eqnarray} 
  \bnabla_\CC G_{opt}^2 = \UUa  ,
	\quad\quad
\bnabla_{\UU_c} G_{opt}^2 =  \Pa \nn + \nnu  \nabla \UUa \bcdot \nn,
	\label{eq:sens_G}
\end{eqnarray} 
where the adjoint base flow $\QQa=(\UUa,\Pa)^T$ is solution of the  linear, non-homogeneous system of equations
\begin{eqnarray} 
& 	\bnabla \bcdot \UUa  =0,
 	\quad
  	-\bnabla \UUa \bcdot  \UU_b + \bnabla \UU_b^T \bcdot  \UUa
	- \bnabla \Pa - \nnu \bnabla^2 \UUa = \bnabla_\UU G_{opt}^2  \quad \mbox{ in } \Omega, \nonumber \\  
	& \UUa = \00  \quad \mbox{ on } \Gamma_w,
\label{eq:adjBF}
\end{eqnarray} 
and 
$\bnabla_\UU G_{opt}^2 =  2 G_{opt}^2 \mbox{Re} (- \bnabla \uu_{opt}^H \bcdot \ff_{opt}
+ \bnabla \ff_{opt}  \bcdot \uu_{opt}^*)$
is the sensitivity to base flow modification,
when the forcing is normalized as
$||\ff_{opt}|| = 1$.
Note that the expression for $\bnabla_\UU G_{opt}^2$  assumes arbitrary base flow variation. 
As mentioned by Brandt \etal \cite{Bra11}, it is possible to restrict this sensitivity field to divergence-free base flow modifications by solving a subsequent  Poisson equation.

For each frequency $\omega$ of interest, 
the optimal forcing and response are computed according to the method described in section \ref{sec:optgain}, 
and  the sensitivity to base flow modification $\bnabla_\UU G_{opt}^2$ is calculated.
Then, the sensitivities to control are obtained as follows:
first, the variational formulation of (\ref{eq:adjBF}) is discretized and solved using \textit{FreeFem++} (with the same mesh and elements as for base flow calculation);
second, sensitivities (\ref{eq:sens_G}) are evaluated. 
The boundary conditions used to compute the adjoint base flow are $\UUa=\00$ at the inlet and on the wall,
$\partial_y  \Ua=\Va=0$ at the top border,
and $\Pa\nn+\nnu\bnabla\UUa\bcdot\nn+\UUa(\UU_b\bcdot\nn)=\00$ at the outlet.

\begin{figure}
 \centerline{
 \begin{overpic}[width=13.3 cm, tics=10]{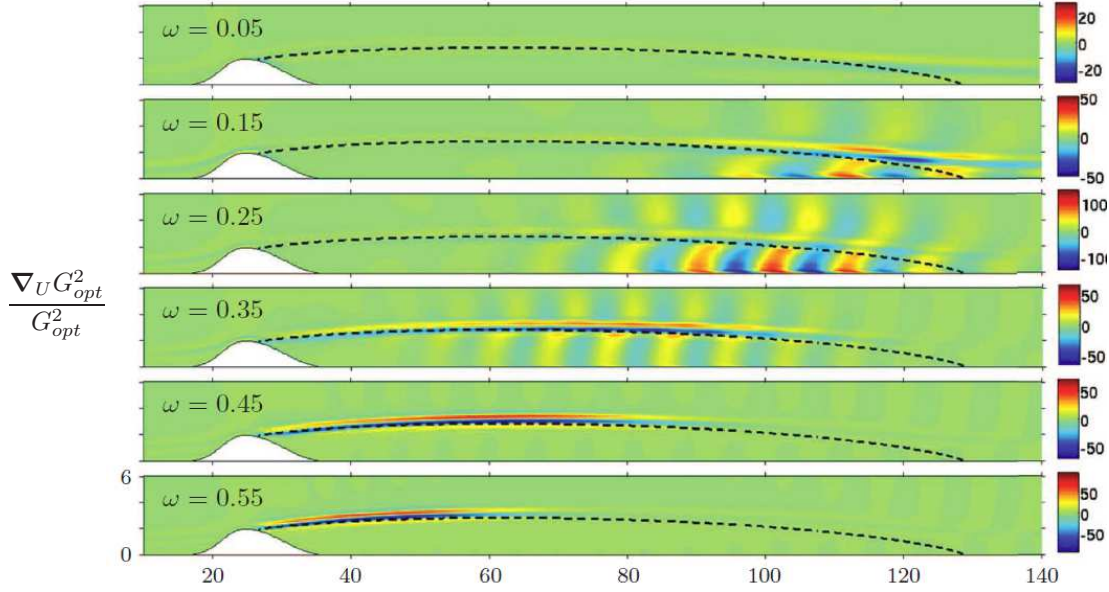}
 	\put(-10, 28)  {\footnotesize $\displaystyle{\frac{\bnabla_{U} G_{opt}^2}{G_{opt}^2}}$} 
 \end{overpic}
 }
  \caption{\small
Normalized sensitivity of optimal gain to base flow modification in the streamwise direction, 
$\bnabla_{U} G_{opt}^2/G_{opt}^2$, at $\Rey=580$ and frequencies $\omega=0.05, 0.15, \ldots 0.55$. The vertical dashed line is the base flow separatrix.
The axes are not to scale.
   }
   \label{fig:optgain_sensit_BF}
\end{figure}

Figure \ref{fig:optgain_sensit_BF} displays the streamwise component of the sensitivity of $G_{opt}^2$ to base flow modification, denoted as  
$\bnabla_{U} G_{opt}^2 = 
 \bnabla_{\UU} G_{opt}^2 \bcdot \ex$,  at $\Rey=580$.
 It shows where a modification of the base flow  $\bdelta\UU=(\delta U,0)^T$  has the largest effect on the optimal gain at each frequency, and if $G_{opt}$ would increase or decrease, according to 
 $\delta G_{opt}^2 = (\bnabla_{\UU} G_{opt}^2  | \bdelta\UU)$.
Two elongated regions of large sensitivity are located in the shear layer and move upstream with $\omega$: a region of negative sensitivity along the separatrix, and a region of positive sensitivity just above. 
The interior of the recirculation region  is  sensitive too at intermediate (i.e. most amplified) frequencies, with structures reminiscent of the optimal response (figure \ref{fig:optforc-optresp}$(b)$).

\begin{figure}
  \centerline{
    \hspace{0.7cm}
  	\begin{overpic}[width=13.25 cm, tics=10]{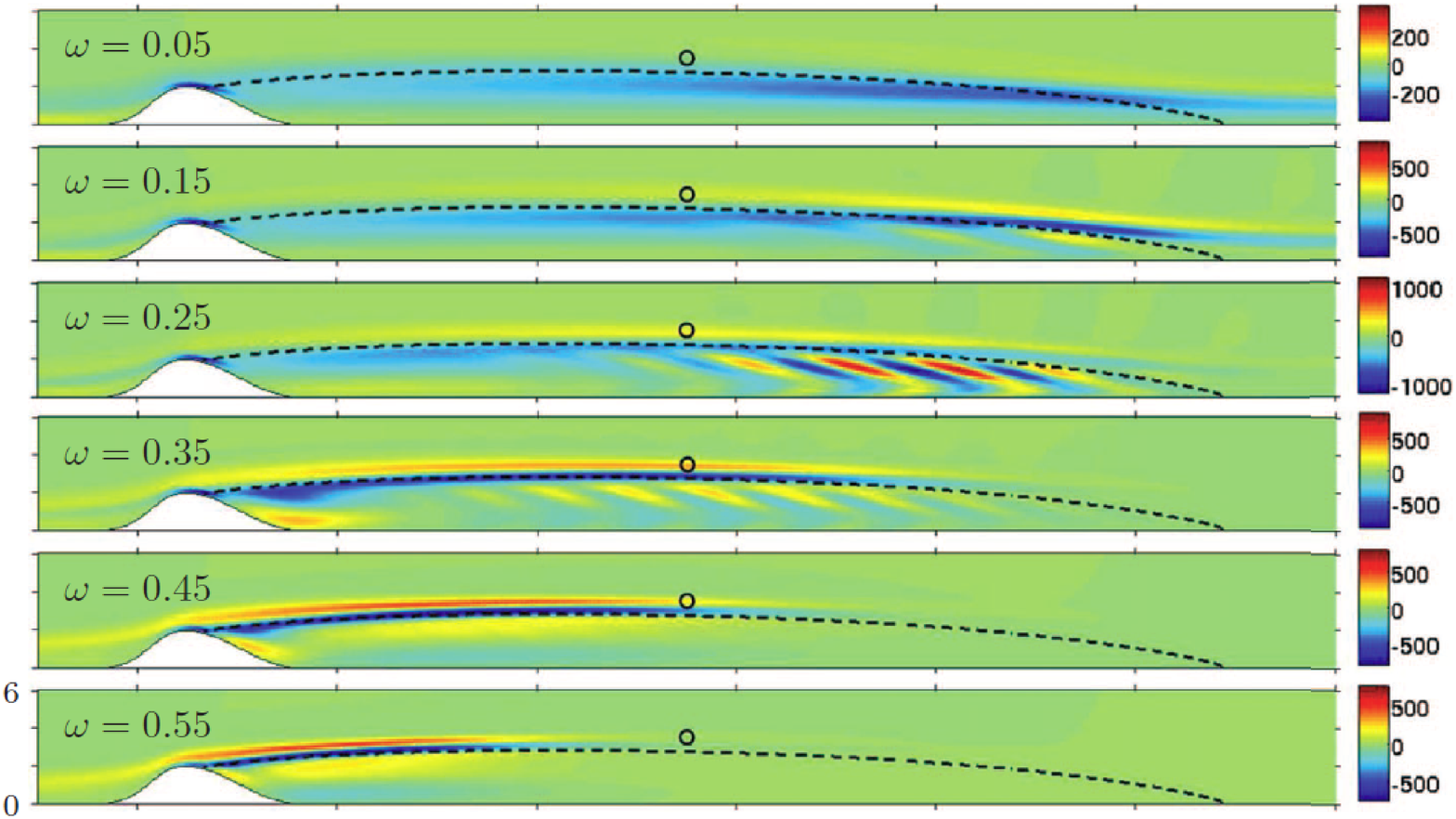}
  	\put(-7.5, 55) {$(a)$}
  	\put(-15, 27.5){\footnotesize $\displaystyle{\frac{\bnabla_{C_x} G_{opt}^2}{G_{opt}^2}}$} 	
  	\end{overpic}
  }
  \vspace{0.4 cm}
  \centerline{
  	\psfrag{x}[t][][1][0]{$x$}
  	\hspace{0.05cm}
  	\begin{overpic}[width=12.3 cm, tics=10]{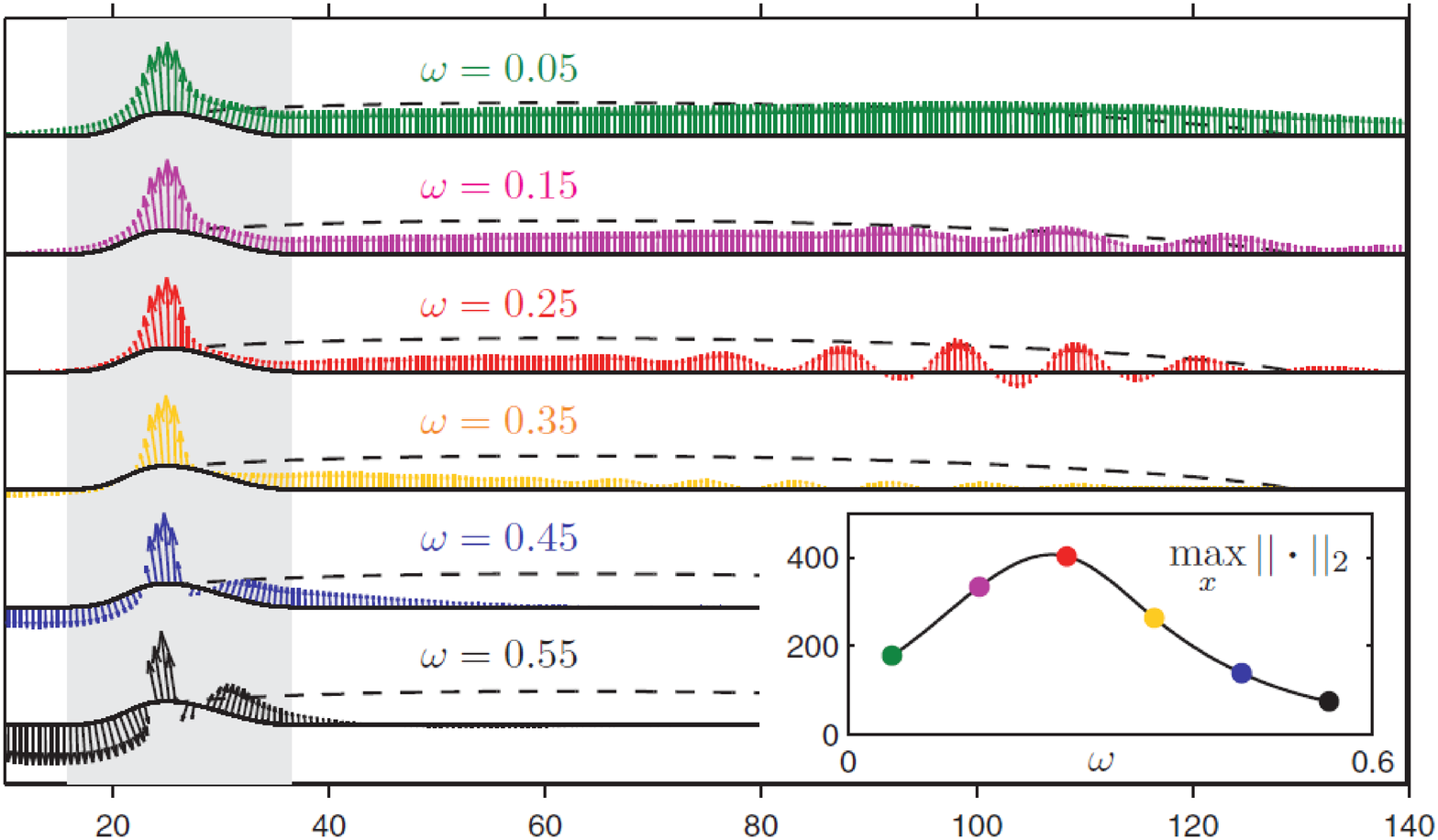}
  	\put(-7., 57) {$(b)$}
   	\put(-15, 31){\footnotesize $\displaystyle{\frac{\bnabla_{\UU_c} G_{opt}^2}{G_{opt}^2}}$} 	
  	\put(48, -2){$x$} 
  	\end{overpic}
  }
  \caption{\small
Sensitivity of optimal gain to control at $\Rey=580$ and frequencies $\omega=0.05, 0.15, \ldots 0.55$.
$(a)$ Normalized streamwise component of the sensitivity to volume control, $\bnabla_{C_x} G_{opt}^2/G_{opt}^2$.
Black circles indicate the location of volume control $(x,y)=(75,3.5)$ discussed in the text and in figure   \ref{fig:optgain_valid_bulk}. The axes are not to scale.
$(b)$~Normalized sensitivity to wall control, $\bnabla_{\UU_c} G_{opt}^2/G_{opt}^2$, rescaled for each frequency by the largest point-wise $L^2$ norm on the wall
$\displaystyle \max_{x} || \bnabla_{\UU_c} G_{opt}^2 /G_{opt}^2||_2$.
This maximal value is shown by symbols in the inset (where the solid line is an indicative fit through the data).
The grey region shows the streamwise extension of the bump. 
The dashed line is the base flow separatrix.
   }
   \label{fig:optgain_sensit_forc}
\end{figure}

Figure \ref{fig:optgain_sensit_forc}$(a)$ shows the streamwise component of the sensitivity of $G_{opt}^2$ to volume control, 
denoted as 
$\bnabla_{C_x} G_{opt}^2 = 
 \bnabla_{\CC} G_{opt}^2 \bcdot \ex$, 
 at $\Rey=580$.
The optimal gain is the most sensitive to control in the shear layer.
However, several observations make difficult the design of an efficient and robust open-loop control based on steady volume control.
First, the location of largest sensitivity (in absolute value) depends on $\omega$: it is close to the reattachment point at lower frequencies, and moves upstream  as frequency increases. Thus, control applied at the location of maximal sensitivity at one frequency will not be optimal at other frequencies.
Second, the sign of the sensitivity depends on space: thin regions of opposite sign are located close to each other (\eg vertically in the shear layer and, for intermediate $\omega$, horizontally in the recirculation region). Slightly misplacing a force intended to reduce the optimal gain might actually increase it.  
Third, in some locations the sign of the sensitivity is varying with frequency.
Therefore, without choosing its location carefully, a control might reduce the optimal gain at some frequencies and increase it at others.
Despite these limitations, one can focus on most amplified frequencies and find a location where volume control reduces the optimal gain at these frequencies. 
At $(x,y)=(75,3.5)$ for instance 
(black circles in figure \ref{fig:optgain_sensit_forc}$(a)$), 
the sensitivity $\bnabla_{C_x} G_{opt}^2$ is positive in the range $0.15\leq\omega\leq0.45$, and small for frequencies outside this range. 
A force located at this location and oriented along $-\ex$ should therefore have an  overall  reducing effect on noise amplification. We will come back to this control configuration later on.

We now look  at the sensitivity of optimal gain to wall control. 
Figure \ref{fig:optgain_sensit_forc}$(b)$ shows  the normalized sensitivity to wall control  $\bnabla_{\UU_c} G_{opt}^2/G_{opt}^2$.
Arrows show the orientation of positive sensitivity, i.e. wall control in the same direction and orientation as the arrows would increase the optimal gain. The sensitivity is essentially normal to the wall, indicating that normal actuation has a much stronger effect than tangential actuation (more specifically: one to two orders of magnitude). 
The sensitivity appears to be maximal at the summit of the bump for all frequencies. 
The maximum point-wise $L^2$ norm along the wall (inset in figure \ref{fig:optgain_sensit_forc}$(b)$)  follows with $\omega$ the same trend as $G_{opt}$, meaning that wall control authority is larger at frequencies which are more amplified.
In addition, one can observe that $\bnabla_{\UU_c} G_{opt}^2$ changes sign 
with $\omega$ at some locations (\eg upstream of the bump, and in the downstream half of the recirculation region); however, at the bump summit it is oriented towards the fluid domain for all frequencies, and therefore   wall suction at this location would reduce $G_{opt}$  for all frequencies.

The above considerations on the sensitivity to volume control and wall control suggest designing the following open-loop control:
no actuation in the domain, $\CC=\00$, and vertical wall suction $\UU_c=(0,U_c)^T$ at the bump summit $x=x_b$.
In the following, the Gaussian profile $U_c(x) = W \exp(-(x-x_b)^2/\sigma_c^2)/(\sigma_c \sqrt{\pi})$, with 2D flow rate $W$, will be applied at the wall $(x,y_b(x))$.

\begin{figure}[b]
  \centerline{ 
    \hspace{0.3 cm}
  	\psfrag{G2}[r][][1][-90]{$G_{opt}^2$}
	\psfrag{dF}[t][][1][0]{$C_x$}
	\psfrag{optimal gain}[r][][1][-90]{$G_{opt}$}
  	\psfrag{omega}[t][][1][0]{$\omega$}
   \begin{overpic}[height=6.2 cm, tics=10]{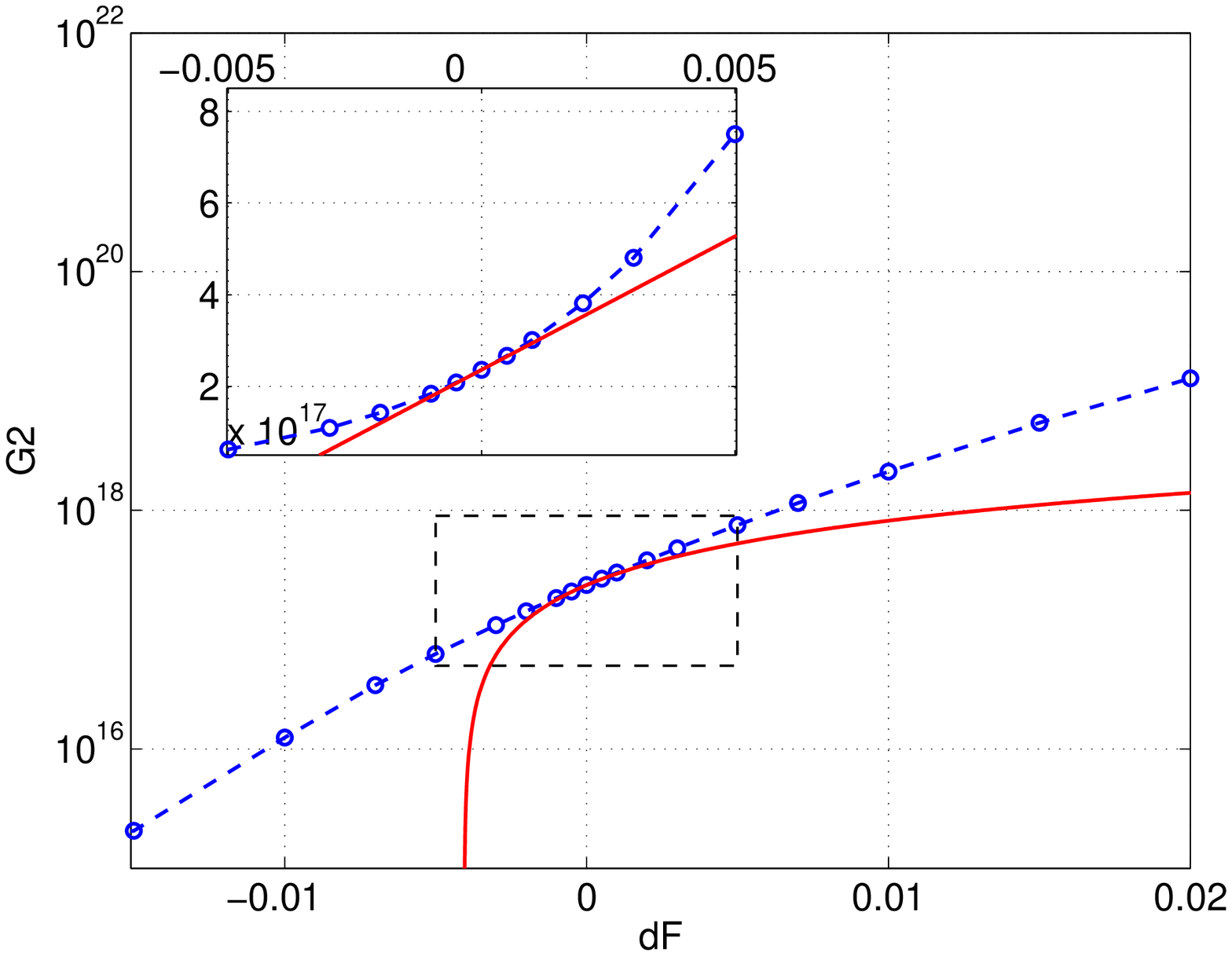}
  	\put(-7, 71) { $(a)$}
  	\put(82, 47) {\footnotesize \textcolor{blue}{NL}}
  	\put(82, 31) {\footnotesize \textcolor{red}{SA}}
  \end{overpic}
   \hspace{0.5 cm}
  \begin{overpic}[height=6 cm, tics=10]{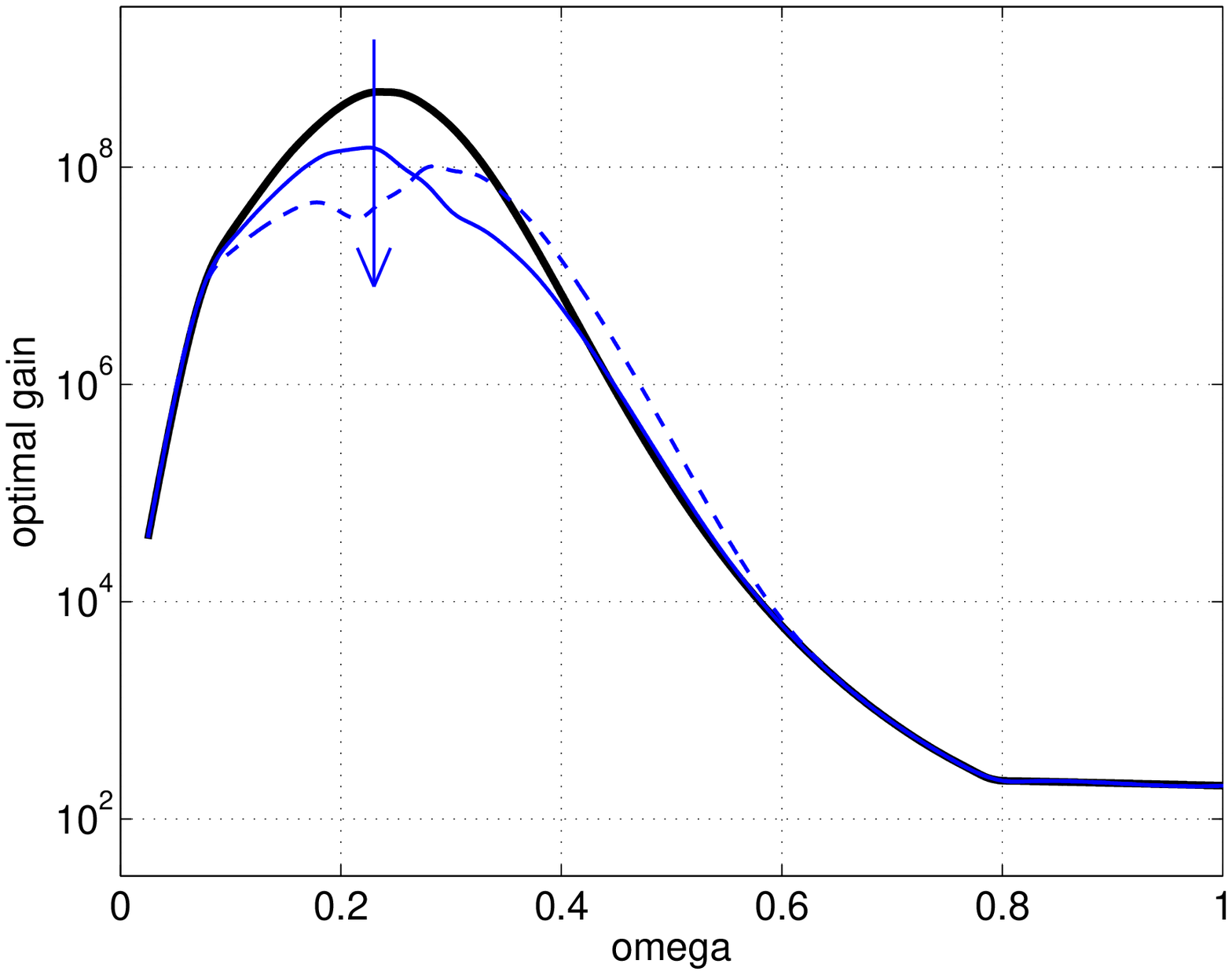}
  	\put(-7, 73) { $(b)$}
   	\put(36,70){                 \footnotesize $C_x=0$ (uncontrolled)}
 	\put(19,46)  {\textcolor{blue}{\footnotesize $C_x=-0.01,$}}
	\put(31,40.5){\textcolor{blue}{\footnotesize     $-0.02$}}
  \end{overpic}
  }
  \caption{
  Variation of the optimal gain at $\Rey=580$ when applying at $(x,y)=(75,3.5)$ a steady volume control  of amplitude $C_x$ in the streamwise direction.
$(a)$
Prediction from sensitivity analysis (SA, red solid line) 
and  non-linear controlled base flows (NL, blue symbols) at $\omega=0.25$.
The main plot is in logarithmic scale,
the inset in linear scale (the sensitivity is a straight line). 
$(b)$
 $G_{opt}(\omega)$ for
$C_x=0$  (thick solid line),
$C_x=-0.01$ (thin solid line) and
$C_x=-0.02$ (dashed line).
}
   \label{fig:optgain_valid_bulk}
\end{figure}

\begin{figure}[b]
  \centerline{
    \hspace{0.3 cm}
  	\psfrag{G2}[r][][1][-90]{$G_{opt}^2$}
	\psfrag{W}[t][][1][0]{$W$}
 	\psfrag{suction}[][][1][0]{\footnotesize suction}
  	\psfrag{blowing}[][][1][0]{\footnotesize blowing}
  	\psfrag{optimal gain}[r][][1][-90]{$G_{opt}$}
  	\psfrag{omega}[t][][1][0]{$\omega$}
  \begin{overpic}[height=6.2 cm, tics=10]{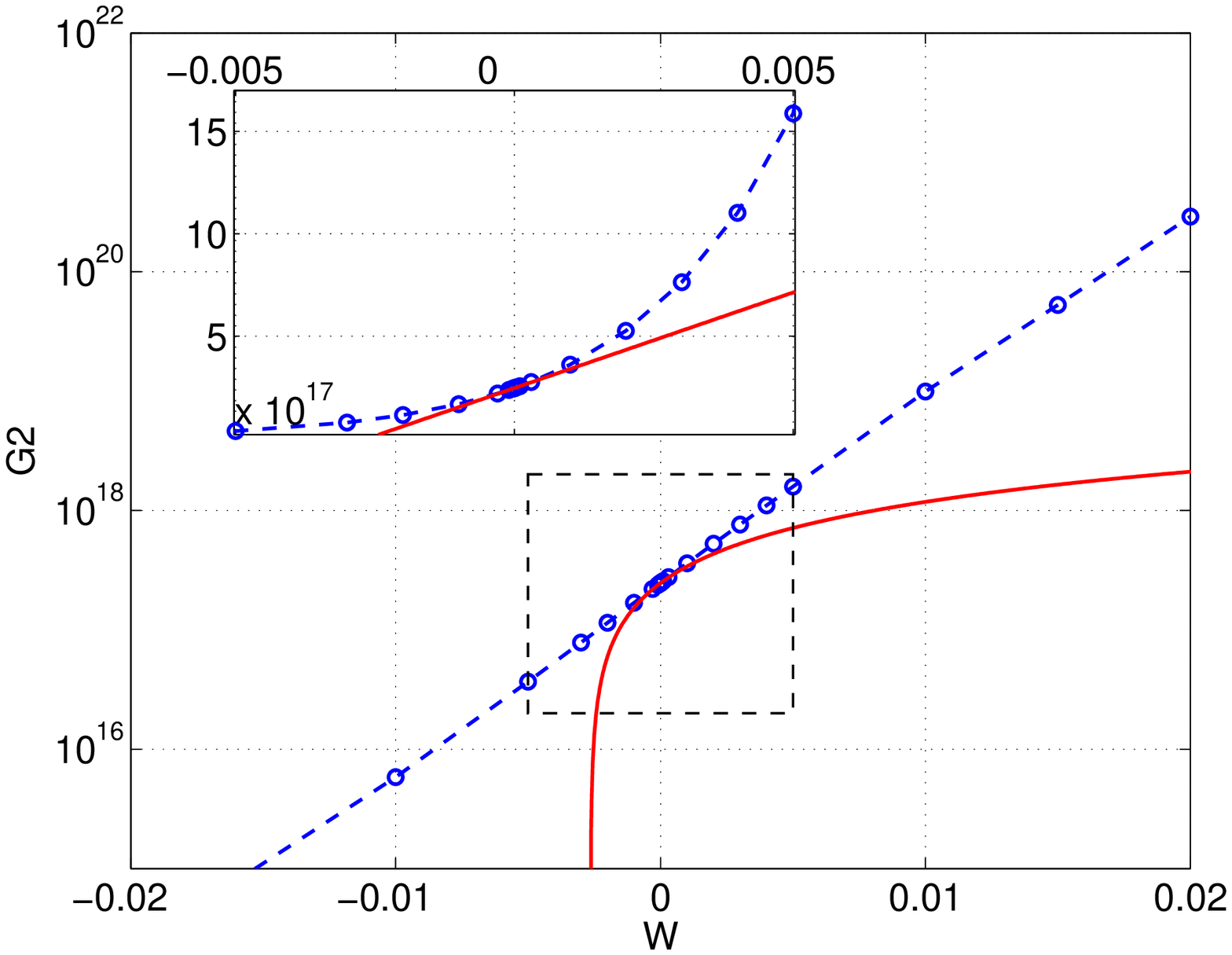}
  	\put(-7, 71) { $(a)$}
  	\put(82, 57) {\footnotesize \textcolor{blue}{NL}}
  	\put(82, 33) {\footnotesize \textcolor{red}{SA}}
  	\put(20, -0.5) {\footnotesize $\longleftarrow$ suction}
  	\put(65, -0.8) {\footnotesize blowing $\longrightarrow$}
  \end{overpic}
  \hspace{0.5 cm}
  \begin{overpic}[height=6 cm, tics=10]{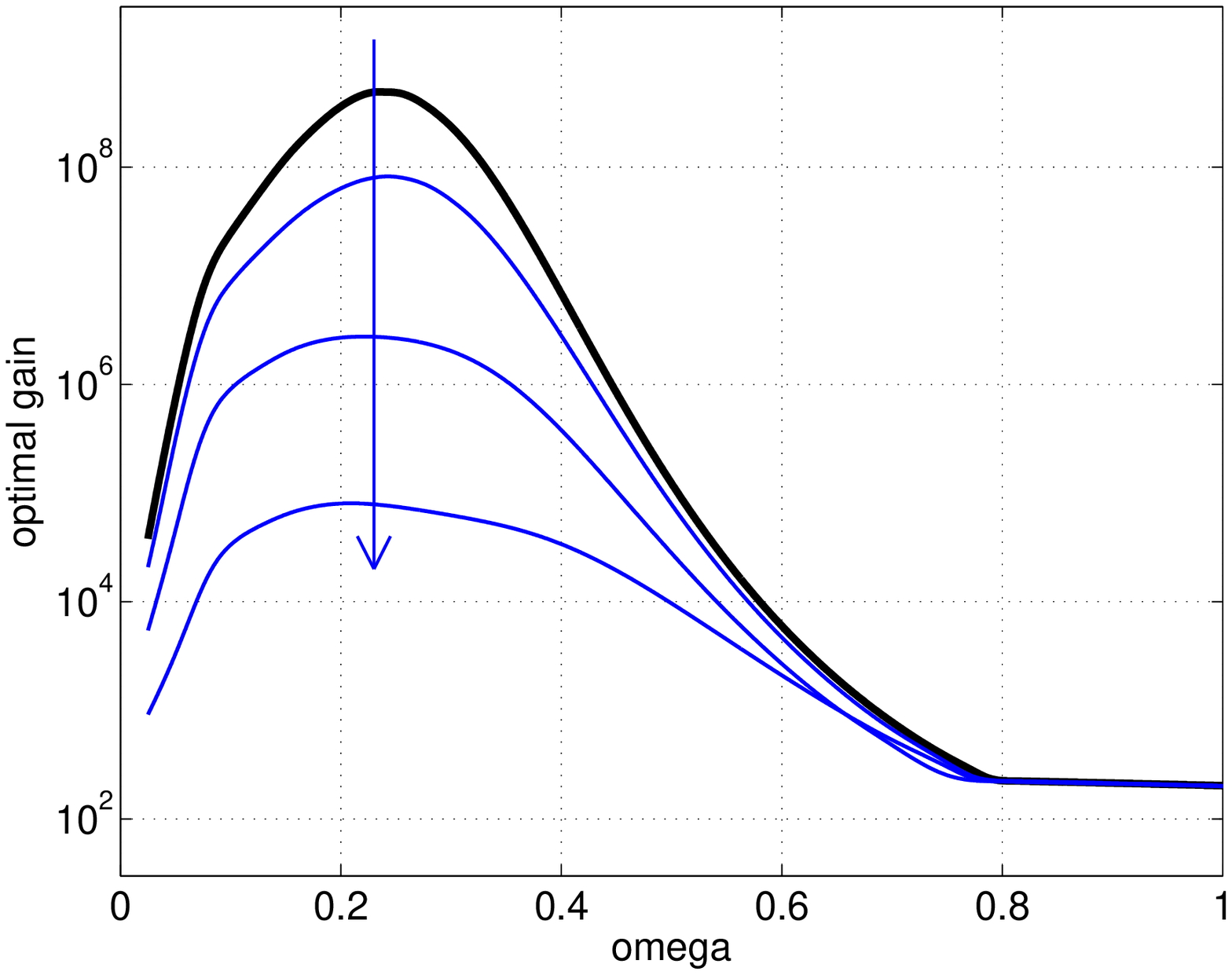}
  	\put(-7,73){ $(b)$}
  	\put(36,70){                 \footnotesize $W=0$ (uncontrolled)}
 	\put(21,23)  {\textcolor{blue}{\footnotesize $W=-0.010,$}}
 	\put(32,17.5){\textcolor{blue}{\footnotesize   $-0.035,$}}
 	\put(32,12)  {\textcolor{blue}{\footnotesize   $-0.100$}}
  \end{overpic}
  } 
  \caption{
  Variation of the optimal gain at $\Rey=580$ when applying vertical wall blowing/suction at the bump summit.
$(a)$
Prediction from sensitivity analysis (SA, red solid line) and  non-linear controlled base flows (NL, blue symbols).
The main plot is in logarithmic scale and  shows that $G_{opt}^2$ varies exponentially with flow rate.
In linear scale (inset), the sensitivity is a straight line. 
$(b)$
Reduction of $G_{opt}(\omega)$  with flow rates  $W=-0.010$, -0.035, -0.100. 
}
   \label{fig:optgain_valid_suction}
\end{figure}

In order to validate the sensitivity calculations, comparisons were made for several volume and wall control configurations.
We present results for two particular configurations in figures  \ref{fig:optgain_valid_bulk} and \ref{fig:optgain_valid_suction}.
First, figure \ref{fig:optgain_valid_bulk}$(a)$ shows the variation of the optimal gain at $\omega=0.25$ when a steady volume force in the streamwise direction $\CC=(C_x,0)^T$ is applied in the flow at $(x,y)=(75,3.5)$. 
Predictions from linear sensitivity analysis 
(with $\delta G_{opt}^2 = ( \bnabla_{\CC} G_{opt}^2 | \bdelta \CC )$)
are compared to calculations of $G_{opt}$ on non-linear controlled base flows.
The agreement is excellent for the slope, with the sensitivity prediction (solid line) tangent to the curve for actual base flows (dashed line) at zero-amplitude control.
However, strong non-linear effects are at hand, with the difference between the two curves quickly growing with $|C_x|$. 
Figure \ref{fig:optgain_valid_bulk}$(b)$ shows the actual optimal gain for different control amplitudes. 
At $C_x=-0.01$, the optimal gain is reduced for frequencies $0.1<\omega<0.4$.
At $C_x=-0.02$ (dashed line), further reduction is obtained for $0.1<\omega<0.25$ 
but not for higher frequencies as strong non-linear effects come into play; 
compared to the uncontrolled case, an increase of $G_{opt}$  is observed for $\omega \geq 0.35$.
Note that the effect of a small control cylinder placed in the flow as in  the experiment of Strykowski and Sreenivasan \cite{Stry90} can be modelled by a volume force of opposite direction and same amplitude as the drag force felt by the control cylinder \cite{Hill92AIAA, mar08cyl, Mel10}. Here, at $(x,y)=(75,3.5)$, the flow is in the streamwise direction (1\% of cross-stream velocity), and given the velocity magnitude, a volume control of $C_x=-0.01$ would correspond to a control cylinder diameter $d=0.007$.

Figure \ref{fig:optgain_valid_suction}$(a)$ shows the variation of the optimal gain at $\omega=0.25$ when vertical wall actuation (blowing/suction) is applied at the bump summit. 
Predictions from linear sensitivity analysis 
(with $\delta G_{opt}^2 = \langle \bnabla_{\UU_c} G_{opt}^2 | \bdelta \UU_c \rangle$)
are compared to calculations of $G_{opt}$ on non-linear controlled base flows 
(with wall blowing/suction actually modelled as a boundary condition).
It appears that $G_{opt}$ varies exponentially with $W$ (straight line in logarithmic scale).
Again, the agreement is very good, and non-linear effects  strong.
Therefore, sensitivity analysis proves useful in identifying sensitive regions to design efficient control configurations, but the final quantitative control performance can only be assessed with non-linear simulations or experiments.
Figure \ref{fig:optgain_valid_suction}$(b)$ shows the actual optimal gain for different suction flow rates, and confirms the efficiency of this control strategy: reasonably small control flow rates achieve a dramatic reduction of $G_{opt}$ for all frequencies, thereby potentially increasing the level of noise the flow can withstand without being destabilized.

\subsection{Reduction of non-linear noise amplification}
\label{sec:DNS_verif}

\begin{figure}[b]
  \centerline{
   	\psfrag{G}[r][][1][-90]{$G_{lin},\,G_{DNS}$}
  	\psfrag{omega}[t][][1][0]{$\omega$}	
  	\psfrag{unc}[][][1][0]{\footnotesize uncontrolled}	
   	\psfrag{W=0}[][][1][0]{\footnotesize $W=0$}	
 	\psfrag{con}[][][1][0]{\footnotesize \textcolor{red}{controlled}}
 	\psfrag{W=0.035}[][][1][0]{\footnotesize \textcolor{red}{$W=-0.035$}}
	\includegraphics[height=6 cm]{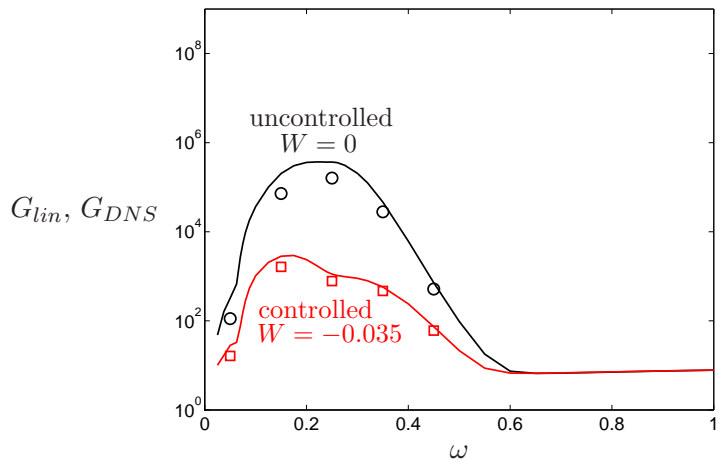}
  }
\caption{
Effect of wall suction on harmonic response.
Upper line and symbols (reported from figure \ref{fig:optgain_DNS}) show the actual gain in the uncontrolled case; 
lower line and symbols are for  wall suction at the bump summit with flow rate $W=-0.035$.
Solid lines:  linear results $G_{lin}$; symbols: $G_{DNS}$  from DNS calculations with small-amplitude harmonic forcing.
   }
   \label{fig:optgain_DNS-ctrl}
\end{figure}

The behavior of the controlled flow is assessed by DNS at $\Rey=580$.
 The same series of simulations as in section \ref{sec:DNS} is performed, 
 now with the steady open-loop control designed in section \ref{sec:sens_opt_gain}, with flow rate $W=-0.035$.
Figure \ref{fig:optgain_DNS-ctrl} compares the actual harmonic gain in the uncontrolled and controlled flows, with the forcing  structure given by (\ref{eq:actuator}).
It shows that the control achieves a significant reduction of about 200 to 300 for the most dangerous frequencies, which are now only amplified by a factor $10^{3}$ instead of $10^{5}$.

\begin{figure}
  \centerline{
   	\psfrag{Em}[r][][1][-90]{$\overline E_p \,\,$}
  	\psfrag{A}[t][][1][0]{$A$}
  	\psfrag{uncontrolled}[][][1][0]{\footnotesize uncontrolled}
  	\psfrag{controlled}[][][1][0]{\footnotesize controlled}
  	\begin{overpic}[height=6 cm, tics=10]{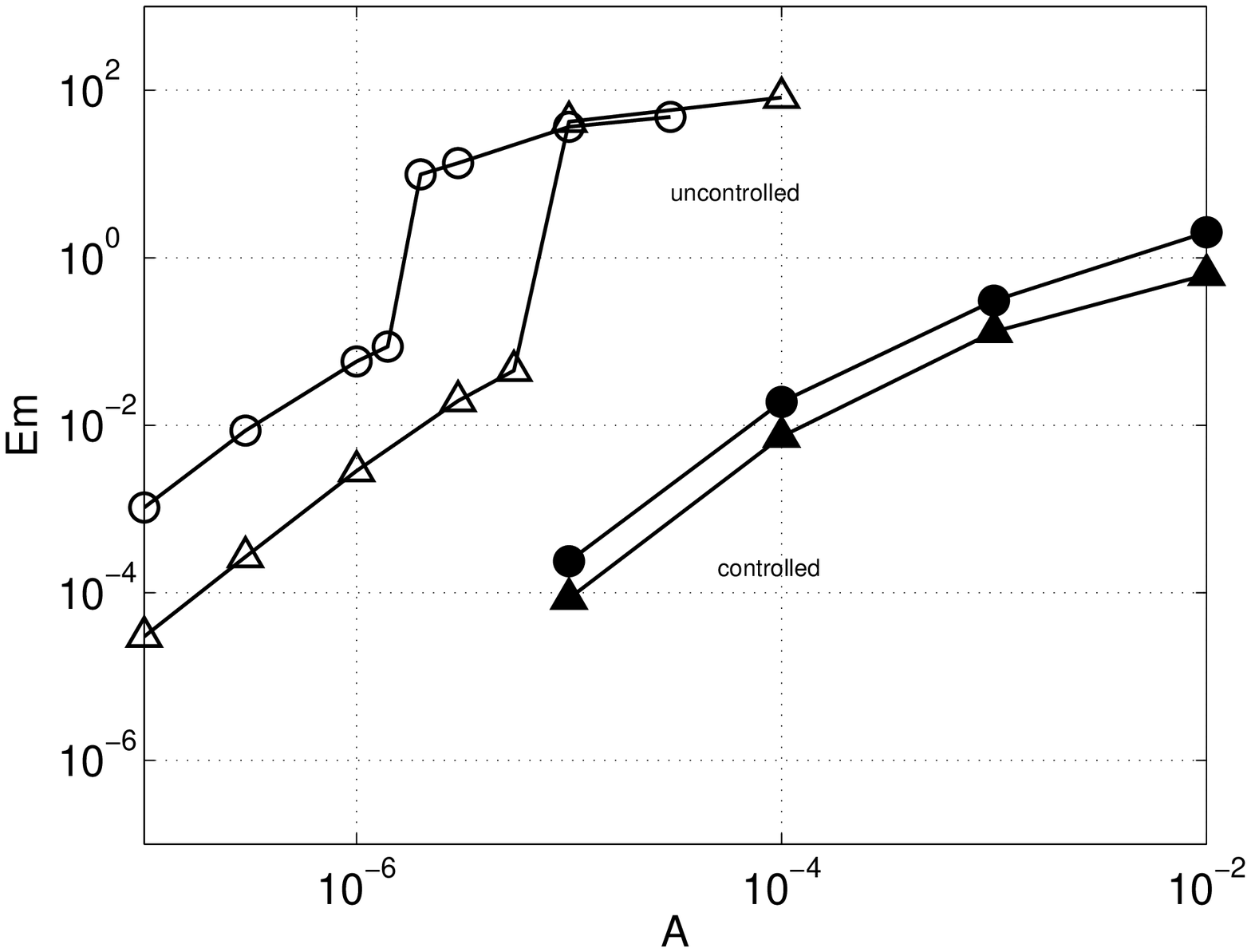}
  	\put(15,69){$(a)$}
  	\end{overpic}
  	\hspace{0.25 cm}
  	\begin{overpic}[height=6 cm, tics=10]{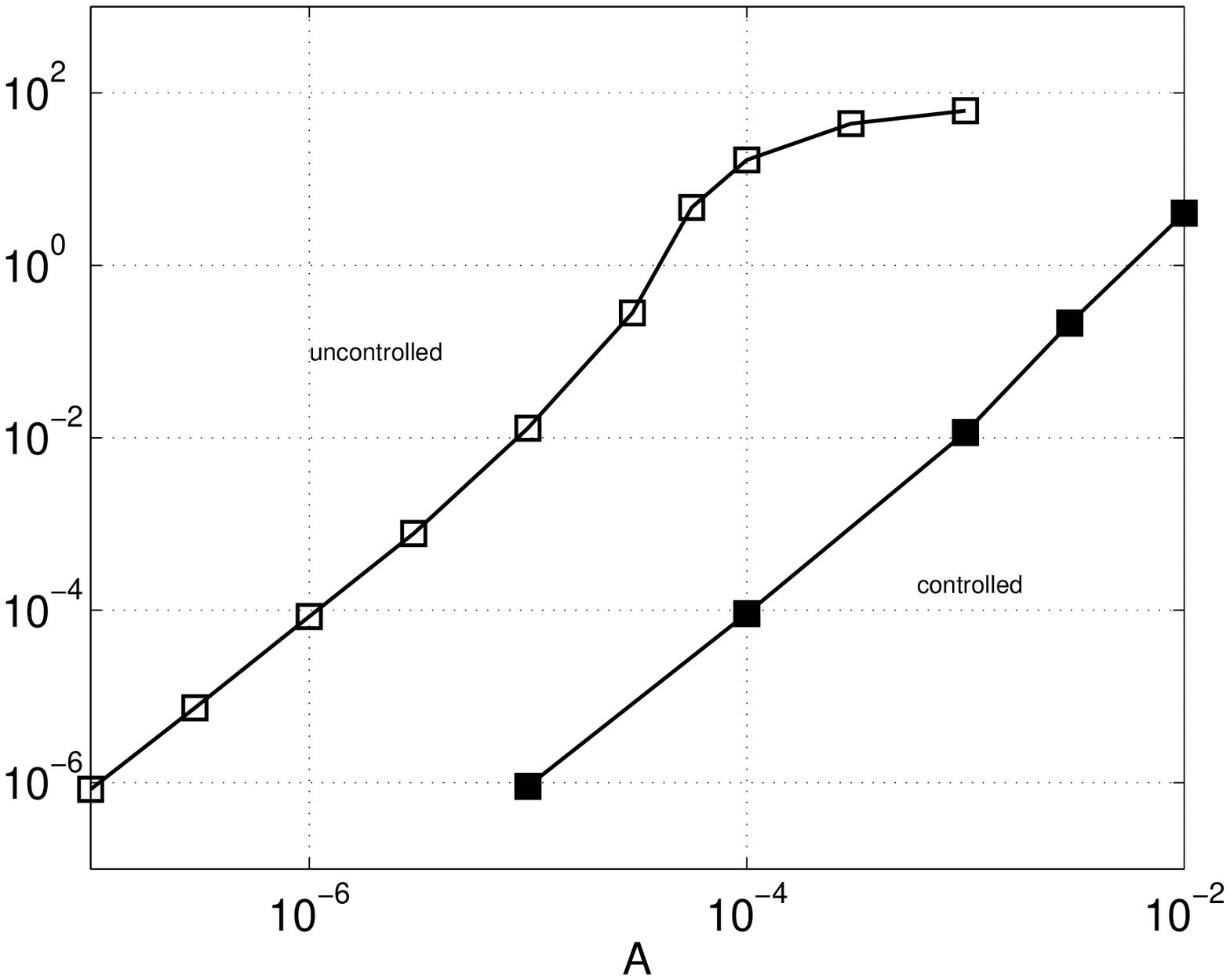}
  	\put(10.5,72){$(b)$}
  	\end{overpic}
  }
  \caption{
Mean asymptotic energy of the perturbations  vs. forcing amplitude, at $\Rey=580$.
Open symbols: without control; 
Filled symbols: with vertical wall suction at the bump summit (flow rate $W=-0.035$).
$(a)$ Harmonic forcing at 
	$\omega=0.25$ (circles) and 
	$\omega=0.35$ (triangles);
$(b)$ stochastic forcing.
   }
   \label{fig:stdE}
\end{figure}

Results for harmonic and stochastic forcing are summarized in figure~\ref{fig:stdE}, which represents the mean asymptotic value of $E_p(t)$.
Typically, amplitudes larger by a factor 100 are needed to reach the same level of energy. 
This is consistent with gain reductions of about 200 to 300 observed 
for the optimal gain in figure 
\ref{fig:optgain_valid_suction}$(b)$
 and, as mentioned above, for the actual gain in figure \ref{fig:optgain_DNS-ctrl}.
The conclusion is the same for harmonic and stochastic forcing: control reduces noise amplification dramatically. The controlled flow can withstand much higher levels of noise than the uncontrolled one 
before being destabilized.

\begin{figure}[h]
  \def\thisfigwidth{7.5 cm}
  \def\thisfigxt{2.5}  
  \def\thisfigyt{3}  
  \centerline{
    \psfrag{E}[r][][1][-90]{$E_p$}
    \psfrag{Ep}[r][][1][-90]{$E_p$}
    \psfrag{U}[][][1][-90]{$U\,\,\,\,$}
  	\psfrag{t}[t][][1][0]{$t$}
  	\begin{overpic}[height=5.8 cm,tics=10]{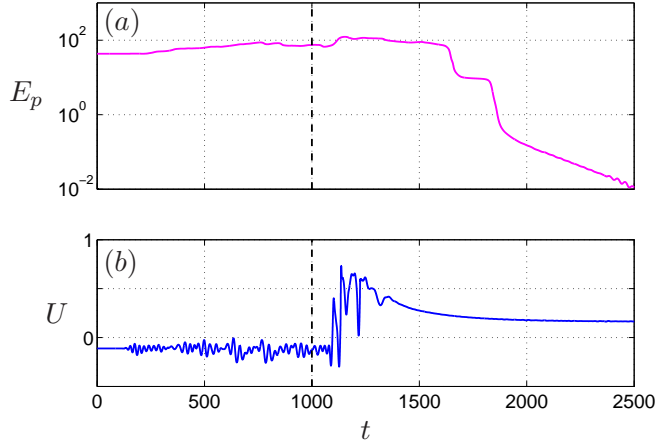}
	\put(12,64){$(a)$}
	\put(12,26){$(b)$}
  	\end{overpic}
  }
  \caption{
Flow restabilization at $\Rey=580$ in direct numerical simulations with steady vertical wall suction at the bump summit (flow rate $W=-0.035$).
$(a)$ Energy of the perturbations (calculated with the final steady-state as reference base flow).
$(b)$ Streamwise velocity of the total flow at $(x,y)=(80,1)$.
The subcritical flow, stationary for $t<0$, is perturbed from $t=0$ with stochastic forcing of amplitude $A=3\times 10^{-4}$, and control is turned on at $t=1000$.
   }
   \label{fig:dns_restab_subcr}
\end{figure}

As an illustration, figure \ref{fig:dns_restab_subcr}
shows the result of a DNS where the flow is forced with stochastic noise of amplitude $A=3\times 10^{-4}$, large enough to destabilize the flow. 
Control with flow rate $W=-0.035$ is  turned on at $t=1000$. The flow is restabilized and becomes stationary. This new steady-state (different from the steady-state at the same Reynolds number without forcing nor control) is used as the reference base flow for the calculation of $E_p$, which quickly drops to zero after control is turned on.
 The streamwise velocity signal  measured at 
$(x,y)=(80,1)$ changes from negative to positive, showing that there is no backflow any more at this location and indicating that wall suction shortens the recirculation region.

\section{Flow stabilization}
\label{sec:stabilization}

\begin{figure}[!h]
  \centerline{
   	\includegraphics[height=10 cm, angle=-90]{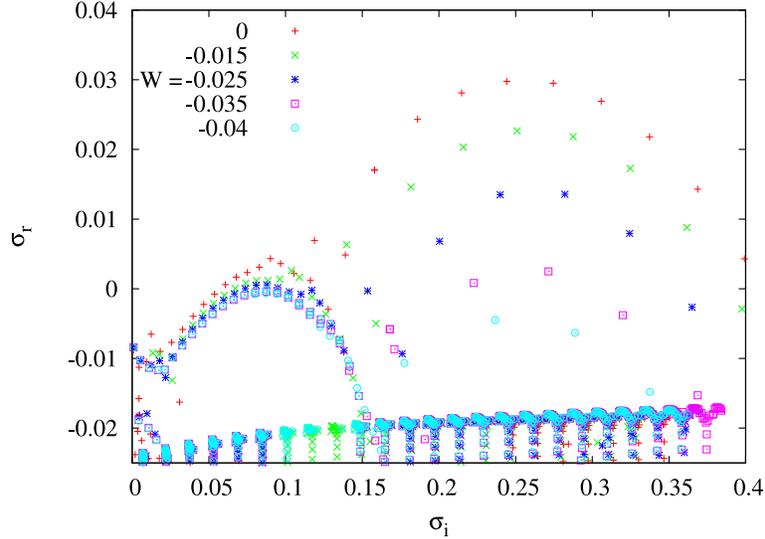}
  }
  \caption{
Global linear eigenspectrum at $\Rey=620$ of the uncontrolled flow and of the flow controlled with vertical wall suction at the bump summit with flow rate $W=-0.015$, -0.025, -0.035, -0.040.
   }
   \label{fig:spectra}
\end{figure}

We  turn our attention to the supercritical regime.
Unlike in the subcritical regime, it is not possible to assess the performance of any control in terms of its effect on optimal gain since 
 the flow is unstable and the notion of asymptotic harmonic response is irrelevant.
The natural tool to be used is global linear stability analysis.
With a global mode  decomposition for perturbations 
$\qq'(x,y,t) = \qq(x,y) e^{\ev t}$, the linearized Navier--Stokes
equations (\ref{eq:LN}) without forcing yield the eigenvalue problem
\be 
	\bnabla \bcdot \uu = 0, 
	\quad
	\ev \uu + \bnabla \uu \bcdot \UU_b + \bnabla \UU_b \bcdot \uu + \bnabla p - \nnu \bnabla^2 \uu = \00,
 \label{eq:EVP}
\ee
where complex  eigenvalues $\ev=\ev_r+i\ev_i$ of positive (resp. negative) real part correspond to unstable (resp. stable) eigenmodes $\qq$.
The aim of the control is now to reduce  the growth rate $\ev_r$ of unstable modes.

The eigenvalue problem (\ref{eq:EVP}) is 
discretized as $(\ev\BB+\LL)\qq=\00$, where
$\LL=\LL(\UU)$ is the Jacobian matrix,
 and
solved at $\Rey=620$ with the method described in Ehrenstein and Gallaire\cite{Ehr08}.
Linearization is first performed around the uncontrolled base flow,  
then around a series of base flows controlled by vertical wall suction at the bump summit with increasing flow rates.
In the uncontrolled case, the flow is globally unstable: we recover the eigenspectrum of Ehrenstein and Gallaire\cite{Ehr08} shown in figure \ref{fig:spectra} and characterized by two distinct branches of eigenvalues.
The most unstable branch corresponds to a family of global modes of similar spatial structure localized around the reattachment point and associated with a Kelvin-Helmholtz instability of the shear layer.
The other branch corresponds to weakly unstable convective modes, typical of  Tollmien-Schlichting instability in boundary layers.
As the control amplitude is increased, both branches become less unstable, until all modes are fully restabilized for $W\simeq -0.040$.
Eigenvalues which are stable in the uncontrolled case are not destabilized by the control.
Therefore the  control strategy designed in section \ref{sec:sens_opt_gain} has a direct and selective effect on unstable eigenvalues, efficiently moving them to the stable half-plane without destabilizing other eigenvalues.
This trend could be expected because the main effect of normal wall suction is to shorten the recirculation region and reduce the strength of the shear layer, thus mitigating shear instabilities.
Since  noise amplification in the subcritical regime and unstable global eigenmodes in the supercritical regime are different manifestations of  the same type of mechanisms (Orr, Tollmien-Schlichting, and more importantly Kelvin-Helmholtz as already mentioned in section \ref{sec:optgain}), it seems reasonable that 
a well-chosen control can have a stabilizing effect on both.

More insight can be gained using a systematic sensitivity analysis to investigate  the effect 
 of steady wall control on most unstable eigenvalues.
Similar to section \ref{sec:sens_opt_gain}
 for the optimal gain,
the variation of a given eigenvalue $\ev$ 
resulting from a small wall actuation $\bdelta \UU_c$ is written as
$\delta \ev = \langle \bnabla_{\UU_c}\ev | \bdelta \UU_c \rangle$. 
Here a discrete method is employed to compute the sensitivity $\bnabla_{\UU_c}\ev$.
The above eigenvalue shift  is equivalent to
$\delta \ev = (\bnabla_{\UU}\ev | \bdelta \UU)$, 
where 
the base flow modification $\bdelta\UU$ caused by wall actuation is solution  of the linear system 
$\LL \bdelta\UU=\bdelta \UU_c$, solved for each wall location,
while the sensitivity $\bnabla_{\UU}\ev$ is computed  once only as 
$\displaystyle \left( \bnabla_{\UU}\LL\qq\right)^H\qq\dag$,
with $\qq\dag$ the adjoint mode associated to the global mode $\qq$.

\begin{figure}[h]
  \def\thisfigwidth{7.5 cm}
  \def\thisfigxt{2.5}  
  \def\thisfigyt{3}  
  \centerline{
    \hspace{1.2cm}
  	\psfrag{x}[t][][1][0]{$x$}
  	\begin{overpic}[height=7.5 cm,tics=10]{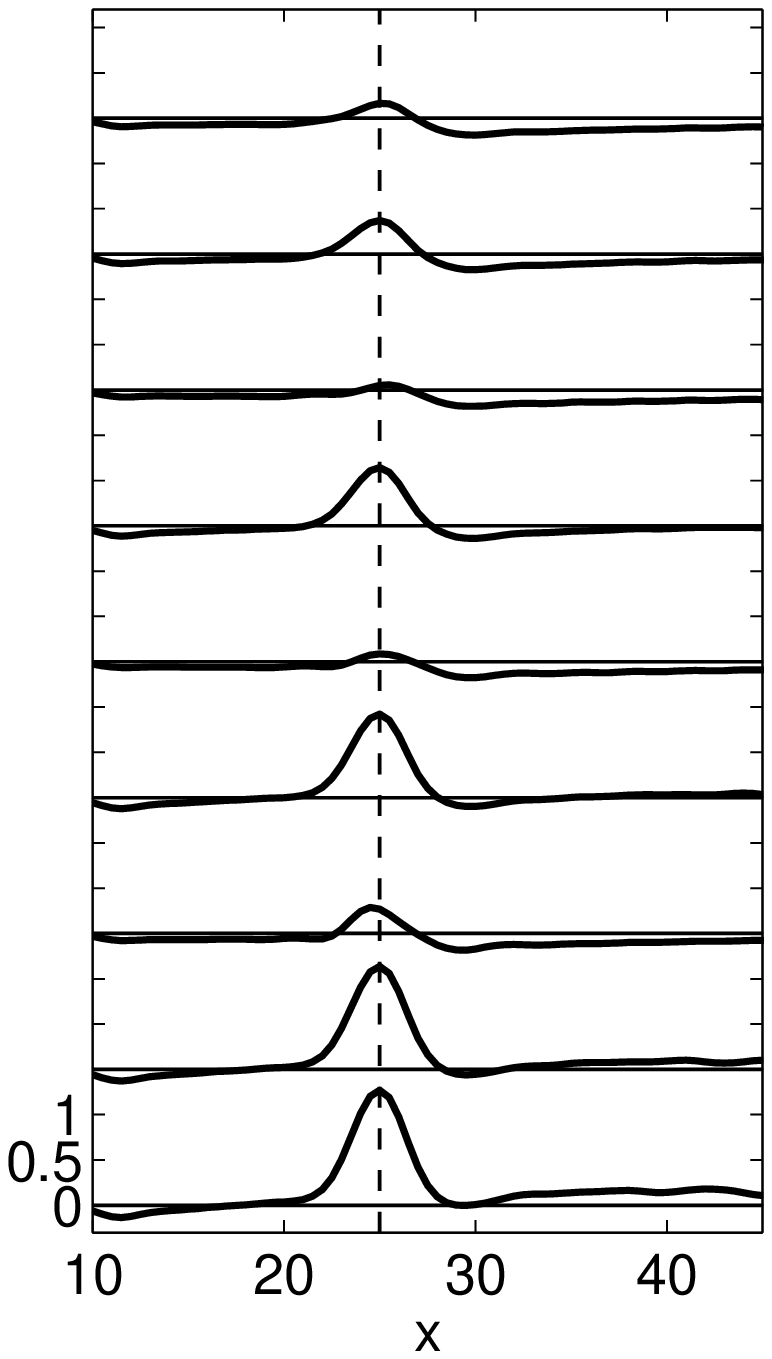}
	  \put(-15,94.5){$(a)$}
	  \put(-21,53){$\bnabla_{\UU_c} \ev_r \bcdot \ey$}
	  \put(37,94)   {\footnotesize mode 1}
	  \put(37,84.25){\footnotesize mode 2}
	  \put(37,74.5) {\footnotesize mode 3}
	  \put(37,64.75){\footnotesize mode 4}
	  \put(37,55)   {\footnotesize mode 5}
	  \put(37,44.25){\footnotesize mode 6}
	  \put(37,35.5) {\footnotesize mode 7}
	  \put(37,25.75){\footnotesize mode 8}
	  \put(37,16)   {\footnotesize mode 9}
	\end{overpic}
    \hspace{0.8cm}
    \psfrag{sr}[r][][1][-90]{$\ev_r \,\,$}
    \psfrag{srr}[r][][1][-90]{$\ev_r \,\,$}
    \psfrag{si}[t][][1][0]{$\ev_i$}
  	\begin{overpic}[height=7.5cm,tics=10]{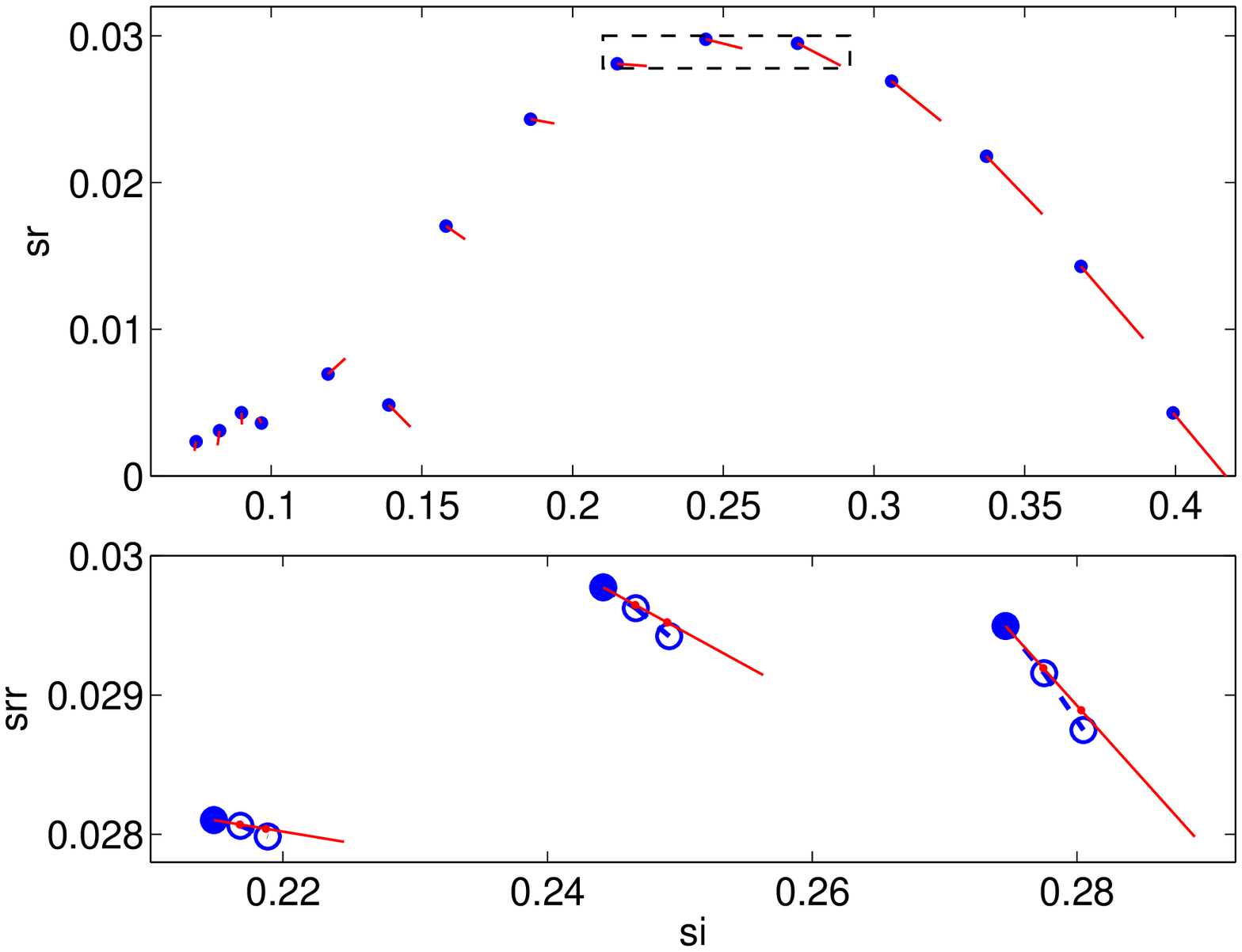}
 	  \put(-3,72){$(b)$}   	 
 	  \put(57,68){\footnotesize 1}
   	  \put(64,67){\footnotesize 2}
 	  \put(51,67){\footnotesize 3}   	  
   	  \put(71,64){\footnotesize 4}  	  
   	  \put(45,63){\footnotesize 5}
   	  \put(77,60){\footnotesize 6}
   	  \put(40,55){\footnotesize 7}   	  
   	  \put(84,52){\footnotesize 8}
   	  \put(91,41){\footnotesize 9}
 	  \put(80,17){\textcolor{blue}{\footnotesize NL}}
 	  \put(93,15){\textcolor{red} {\footnotesize SA}} 
	\end{overpic}
  }
  \caption{
Sensitivity analysis of the most unstable eigenvalues at $\Rey=620$.
$(a)$ 
Sensitivity of the growth rate  of  modes 1 to 9 (Kelvin-Helmholtz branch) to vertical wall control. The dashed line shows the bump summit location.
$(b)$ 
Effect of vertical wall control at the bump summit, as  
predicted by sensitivity analysis. Red solid lines indicate a flow rate $W=-0.005$.
The lower panel is a close-up view of  eigenvalues 1 to 3, comparing sensitivity analysis (SA, red solid lines) and linear stability analysis results for non-linear  base flows controlled with $W=-0.001$ and $-0.002$ (NL, blue circles).
   }
   \label{fig:SA_ev}
\end{figure}

Figure  \ref{fig:SA_ev}$(a)$ shows results for the most unstable eigenmodes of the Kelvin-Helmholtz branch. At the bump summit ($x=25$) the sensitivity of their growth rate to vertical actuation along $\ey$ is positive, therefore vertical wall suction has a stabilizing effect on all these modes.
Any other control configuration would be less effective. For instance, vertical wall blowing at $x=30$ would be slightly more effective in stabilizing modes 3 and 5, but would require more control amplitude, and  might also destabilize mode 9. 
Figure  \ref{fig:SA_ev}$(b)$ shows that convective eigenmodes ($\ev_i \leq 0.15$) are weakly sensitive. 
Global eigenvalues calculated from non-linear  base flows controlled at the bump summit with small-amplitude vertical suction ($W=-0.001$ and $-0.002$) closely follow prediction from sensitivity analysis, as illustrated in the close-up view.

\begin{figure}[h]
  \def\thisfigwidth{7.5 cm}
  \def\thisfigxt{2.5}  
  \def\thisfigyt{3}  
  \centerline{
    \psfrag{E}[r][][1][-90]{$E_p$}
    \psfrag{Ep}[r][][1][-90]{$E_p$}
    \psfrag{U}[][][1][-90]{$U\,\,\,\,$}
  	\psfrag{t}[t][][1][0]{$t$}
  	\begin{overpic}[height=5.8 cm,tics=10]{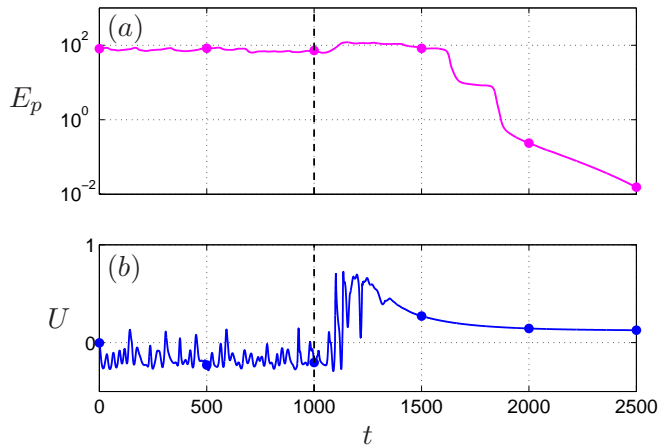}
	\put(12,63){$(a)$}
	\put(12,25.5){$(b)$}
	\end{overpic}  
  }
  \caption{
Flow restabilization at $\Rey=620$ in direct numerical simulations with steady vertical wall suction at the bump summit (flow rate $W=-0.035$).
Same notations as  figure  \ref{fig:dns_restab_subcr}.
The supercritical flow is naturally unsteady, no perturbation is added, and control is turned on at $t=1000$.
Dots correspond to the times of snapshots in figure \ref{fig:dns_restab_field}.
   }
   \label{fig:dns_restab_supercr}
\end{figure}

\begin{figure}[]
  \def\thisfigxt{2.}  
  \def\thisfigyt{3}  
  \centerline{
  	\begin{overpic}[width=15cm,tics=10]{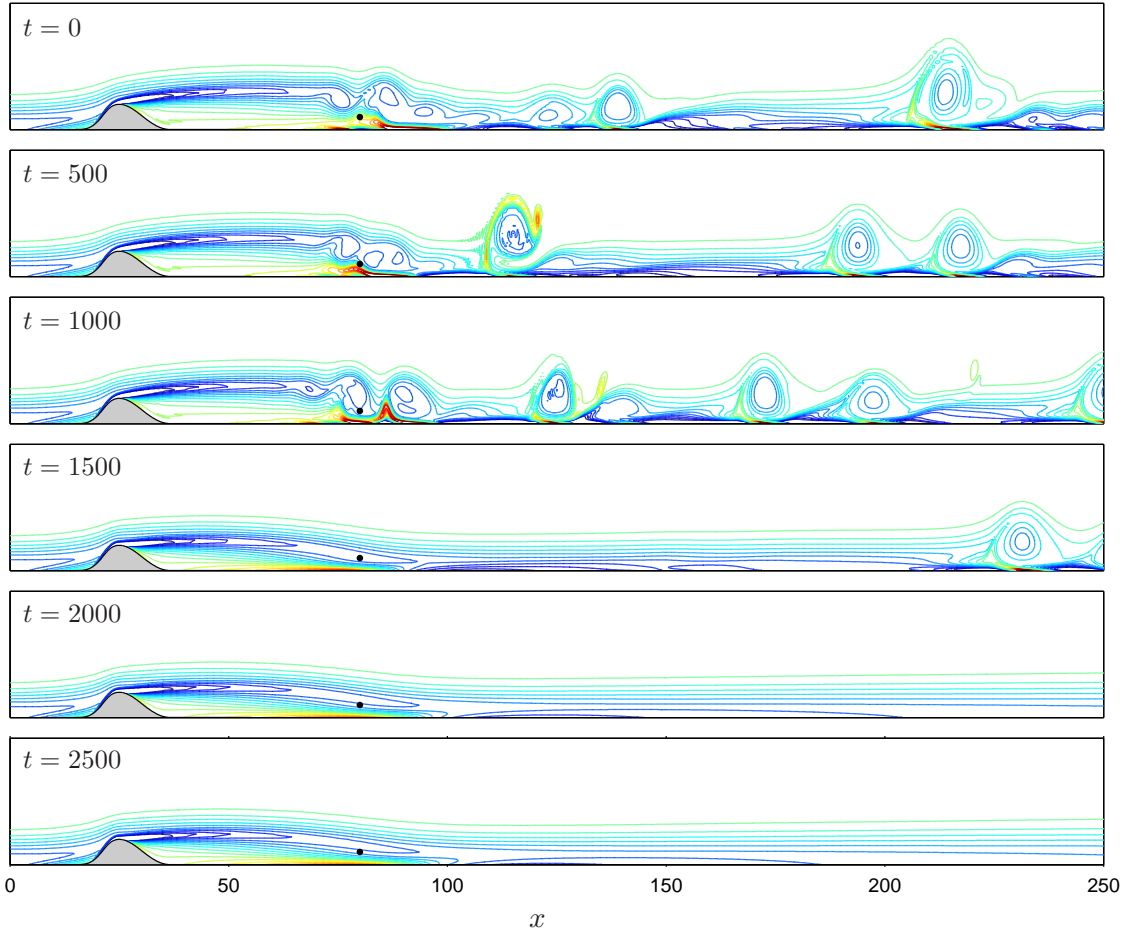}
	\put(\thisfigxt,76)	{\footnotesize $t=0$}
	\put(\thisfigxt,63)	{\footnotesize $t=500$}
	\put(\thisfigxt,50)	{\footnotesize $t=1000$}	
	\put(\thisfigxt,37)	{\footnotesize $t=1500$}	
	\put(\thisfigxt,24)	{\footnotesize $t=2000$}
	\put(\thisfigxt,11) {\footnotesize $t=2500$}
	\put(46,-3){ $x$}	
  	\end{overpic}
  	\vspace{0.3 cm}
   }
  \caption{
Flow restabilization  in the supercritical regime, $\Rey=620$, in DNS 
with steady vertical wall suction at the bump summit (flow rate $W=-0.035$): 
contours of vorticity of the total flow at $t=0$, 500, 1000$\ldots$ 2500. 
The black dot shows the location of the point $(x,y)=(80,1)$ where the velocity signal of figure \ref{fig:dns_restab_supercr} is recorded.
The axes are not to scale.
   }
   \label{fig:dns_restab_field}
\end{figure}

Finally, direct numerical simulations were  performed at several supercritical Reynolds numbers.
Since the flow is naturally unstable, no forcing was added, 
and self-sustained oscillations characterized by low-frequency, large-scale vortex shedding \cite{Mar03} developed. 
Steady vertical wall suction at the summit was turned on at $t=1000$.
Figure \ref{fig:dns_restab_supercr} illustrates how the flow was fully  restabilized at $\Rey=620$ with control amplitude $W=-0.035$.
(The eigenspectra of figure \ref{fig:spectra} suggest that the flow is still unstable with this flow rate. This is due to the different domain size and numerical methods used in the linear stability analysis and in the DNS.)
As in the subcritical case, the streamwise velocity measured at $(x,y)=(80,1)$ is largely fluctuating in the uncontrolled regime, but quickly reaches a steady value once control is turned on. It changes from negative without control to positive with control, because wall suction shortens the recirculation region.
The energy of the perturbations (with the final steady-state taken as reference base flow) quickly decreases to zero as the flow is stabilized and perturbations are advected downstream.
Snapshots of the vorticity field in figure  \ref{fig:dns_restab_field}  clearly
depicts how large-scale perturbations are advected while the control efficiently prevents the formation of new structures and finally drives the flow to a perfectly steady state.

Other direct numerical simulations at $\Rey=620$, which is only slightly supercritical, yielded stable flows with a control amplitude as low as $W=-0.010$, while with $W=-0.035$ the flow could be restabilized for $\Rey \geq 700$.
We did not attempt to determine accurate threshold values of restabilizing control amplitudes $W_c(\Rey)$.

\section{Conclusions}
\label{sec:conclu}

The maximal possible linear amplification of harmonic forcing
was computed at several frequencies in the globally stable 2D separated boundary layer past a wall-mounted bump.
Very large values of the linear optimal gain confirmed the strong non-normal character of this flow, which had already been evidenced by large transient growth in previous studies\cite{Ehr08,Ehr11}.
DNS confirmed that a small-amplitude noise, harmonic or stochastic in time, could lead to a subcritical bifurcation by destabilizing the flow and triggering random unsteadiness.

Using sensitivity analysis, regions where steady control has a desirable reducing effect on optimal gains were identified.
A simple open-loop control inspired by this analysis successfully reduced linear asymptotic response to harmonic forcing at all frequencies.
DNS revealed that this control efficiently dampened noise amplification in the non-linear regime too, which demonstrates that linear analysis captures the essential mechanisms involved in non-modal growth, and  is able to delay bypass transition in such separated open flows.

The success of the present sensitivity-based control method is encouraging.
While being based on physical grounds, it keeps the final design both simple and efficient.
The control strategy,  optimally designed in the subcritical regime, is able 
not only to dampen noise amplification and delay bypass transition in the subcritical regime, 
but also to restabilize the unstable flow in the supercritical regime.

We plan to pursue this study further. First, an ongoing experiment will tell whether this control strategy is robust to 3D effects, and to noise with realistic time and space distributions. Second, it would be useful to circumvent the need to repeat sensitivity analysis 
at each frequency of interest; this will require finding a suitable surrogate  for optimal gain.
\\

This work is supported by the Swiss National Science Foundation (grant no. 200021-130315) and  the French National Research Agency (project no. ANR-09-SYSC-001).

\bibliography{biblio_bump_pof}

\end{document}